\documentclass{ar2e}

\usepackage{amssymb}  
\usepackage{graphicx}

\begin{document}

\input psfig.sty

%
%

\setcounter{topnumber}{2}
\renewcommand\topfraction{1.0}
\setcounter{bottomnumber}{1}
\renewcommand\bottomfraction{.3}
\setcounter{totalnumber}{3}
\renewcommand\textfraction{.0}
\renewcommand\floatpagefraction{.5}
\setcounter{dbltopnumber}{2}
\renewcommand\dbltopfraction{.7}
\renewcommand\dblfloatpagefraction{.5}

\setlength{\baselineskip}{14pt}
\renewcommand\baselinestretch{1}

\setlength{\oddsidemargin}{0.5in}
\setlength{\evensidemargin}{0.3in}

\newcommand{\be}{\begin{equation}}
\newcommand{\ee}{\end{equation}}
\newcommand{\ba}{\begin{eqnarray}}
\newcommand{\ea}{\end{eqnarray}}
\newcommand\simgreater{\buildrel > \over \sim}
\newcommand\simless{\buildrel < \over \sim}
\newcommand{\bt}{\begin{tabular}}
\newcommand{\et}{\end{tabular}}
\newcommand{\half}{{\frac{1}{2}}} 
\newcommand{\MeV}{{\rm MeV}} 
\newcommand{\fm}{{\rm fm}} 


\jname{Annu. Rev. Nucl. Part. Sci.}
\jyear{(2006)}
\jvol{Vol. 56}
\ARinfo{}

\title{\vspace{-5mm}
       DENSE MATTER IN COMPACT STARS: Theoretical Developments and Observational Constraints
       \\ \rule[3mm]{\textwidth}{1mm} \vspace{-15mm} }

\markboth{Page \& Reddy}{Dense Matter in Compact Objects}

\author{
Dany Page
\affiliation{Instituto de Astronom\'{\i}a, 
             Universidad Nacional Autonoma de M\'exico,\\
             Mexico, DF 04510, Mexico;
             email: {\tt page@astroscu.unam.mx}} 
Sanjay Reddy
\affiliation{Theoretical Division, Los Alamos National Laboratory, \\
             Los Alamos, NM 87545, USA;
             email: {\tt reddy@lanl.gov} \vspace{-5mm} }
}

\begin{keywords}
neutron stars, phase transitions, quantum chromodynamics (QCD), superconductivity
\vspace{-5mm}
\end{keywords}

\begin{abstract}
We review theoretical developments in studies of dense matter and its phase structure 
of relevance to  compact stars. 
Observational data on compact stars, which can constrain the properties of dense matter, 
are presented critically  and interpreted.
\end{abstract}

\maketitle

 \section{INTRODUCTION}

Since their discovery almost 40 years ago, neutron stars have been recognized as  
promising laboratories for studying matter under extreme conditions. 
All the ambient conditions that characterize these objects tend to the extreme. 
The typical density inside a neutron star is comparable to the density inside nuclei, 
$\rho_\mathrm{nuclear}=2.5 \times 10^{14}$ g cm$^{-3}$, which corresponds to a number 
density of baryons $n_B \simeq 0.15 $ fm$^{-3}$.  
Under these conditions the nucleons are very degenerate, with a typical Fermi momentum 
$k_F \sim (3 \pi^2 n_B)^{1/3}$ in the range of $300$--$600$ MeV. 
In the interior, the interaction energy of the nucleons is several times larger than their Fermi 
degeneracy energy $E_F = k_F^2/2m \sim 60$--$150$ MeV.  
The baryon chemical potential, which is defined as the energy needed to introduce 
one unit of baryon number, 
can be $500$--$1000$ MeV larger, at the center of the star, than the rest mass of the nucleon. 
There the internucleon distance becomes comparable to their intrinsic size of $1$ fm.  
Under such extreme conditions several new states of matter, containing new high-energy degrees 
of freedom, may be favored. 
Novel hadronic phases -- which contain, in addition to nucleons, a Bose condensate of pions or kaons,
and/or hyperons -- have been proposed and studied in some detail.  
There is also the exciting possibility that the core may contain deconfined quark matter,
much like the quark-gluon plasma being probed in terrestrial experiments such as the 
Relativistic Heavy Ion Collider (RHIC). 
Although model descriptions of these various phases are still in rudimentary form, 
in several cases qualitative differences between the exotic phases and nuclear matter
have been identified. 
A quantitative understanding of these trends and, in particular, how they could affect 
observable aspects of compact star structure and evolution is an exciting area of research. 
A primary purpose of this review is to convey some of this excitement in a pedagogic style 
to make it accessible to graduate students. 
This review is not intended to be a comprehensive account of all recent developments. 
Instead, we have chosen topics that serve (in our view) as pedagogic tools that help 
illustrate key concepts of relevance to dense matter in compact stars. 

\begin{figure}[h]
\begin{center}
\includegraphics[width=0.8\textwidth,angle=-0]{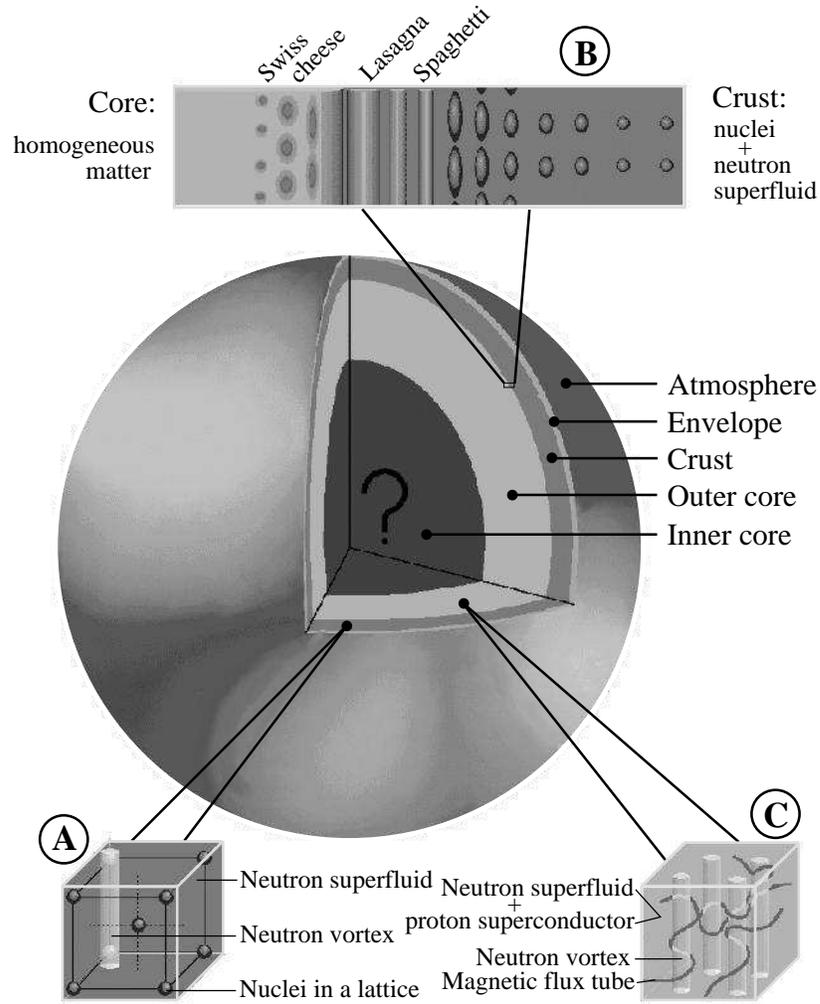}
\caption{Inside a Neutron Star.}
\label{fig:Nstar}   
\end{center}   
\end{figure}

Figure \ref{fig:Nstar} illustrates a theoretician's view of a neutron star. 
The environment just outside the star, characterized by large magnetic (and electric) 
fields and  temperatures, is extreme from plasma physics point of view. 
At the surface of the star we expect a very thin atmosphere composed of hydrogen 
and in some cases perhaps a mix of heavy elements, or even a condensed magnetic surface.
This surface is of utmost observational importance because it allows us to determine the 
temperature, and possibly the radius, of the neutron star, 
an issue we describe in Section~\ref{sec:obs-cooling}.
An envelope exists just below the atmosphere where matter is not yet fully degenerate
and, with a thickness of a few tens of meters, 
it acts as a thermal insulator between the hot interior and the surface.
The outer $500$--$1000$ meters of the star, its crust, contains nuclei, forming a 
lattice immersed in a quantum liquid of neutrons, most probably in a superfluid state 
(Figure~\ref{fig:Nstar}, insert A).
Owing to the rotation of the star, the neutron superfluid forms vortices, which very likely 
pin on the nuclei and participate in the glitches.
As we go deeper into the crust, when approaching nuclear density there is a first-order
phase transition from the inhomogeneous regime of the crust to the homogeneous core.
In this regime, termed the nuclear pasta (Figure~\ref{fig:Nstar}, insert B), 
as density grows, nuclei are increasingly elongated (the spaghetti phase),
then form bidimensional structures (lasagnas) 
-- with the space between them still filled by the superfluid neutrons --
untill the geometry inverts in the Swiss cheese phase, where bubbles of neutrons
are immersed in the almost homogeneous neutron+proton liquid.
The physical reasons for the peculiar aspect of this phase transition are elucidated in 
Section~\ref{sec:pasta}.
At a density of approximately $0.6 \; \rho_\mathrm{nuclear}$, we enter the core 
where the neutron superfluid coexists with the proton fluid, which is probably a Type II 
superconductor that reacts to the neutron star magnetic field by confining it into
fluxoids (Figure~\ref{fig:Nstar}, insert C).
A few kilometers below may be the inner core, marked in Figure~\ref{fig:Nstar} by ``?''.
This question mark is the focus of this review.

In Section~\ref{sec:section2} we present a selected set of observational constraints.
These include mass (Section~\ref{sec:masses}) and radii (Section~\ref{sec:obs-cooling}) measurements 
that probe the equation of state (EoS), i.e., the high-energy properties of dense matter.
Measurements of temperatures and thermal luminosities of isolated neutron stars and accreting 
ones in binary systems are presented in Section~\ref{sec:obs-cooling} and Section~\ref{sec:SXRT},
respectively.
These observations provide information about the structure of the stars, i.e., the EoS, but also
about the low-energy response functions, such as specific heat, transport properties, and neutrino
emission, which probe the low-energy (keV--MeV) excitation spectrum.
In Section~\ref{sec:section3} we discuss the essential physics of dense nuclear matter and 
describe the current state of model calculations of the nuclear EoS and low-energy response. 
In Section~\ref{sec:section4} we describe theoretical models for phase transitions to novel states 
of matter such as Bose-condensed hadronic matter, hyperonic matter, and quark matter at 
supranuclear density. 
Section~\ref{sec:section5} confronts theoretical models of neutron star structure and of evolution 
derived from the studies described in the two previous sections with the observational baggage 
acquired in Section~\ref{sec:section2}. 
Finally, we offer some conclusions in Section~\ref{sec:section6}.

\section{OBSERVATIONAL CONSTRAINTS} 
         \label{sec:section2}

Among the more than 100 billion stars in the Milky Way, almost 1\% of them
($\sim 10^8$--$10^9$) are expected to be compact stars.
Nevertheless, the number of known compact stars is of the order of only a couple
thousand, the majority being radio pulsars (more than 1500) and the rest comprising
approximately 300 accreting stars in binary systems, plus a few dozen single stars
that do not show any pulsed radio emission and were discovered through their 
high-energy emission. 

\subsection{Mass Measurements} \label{sec:masses}

\begin{figure}[t]
\begin{center}
 \psfig{figure=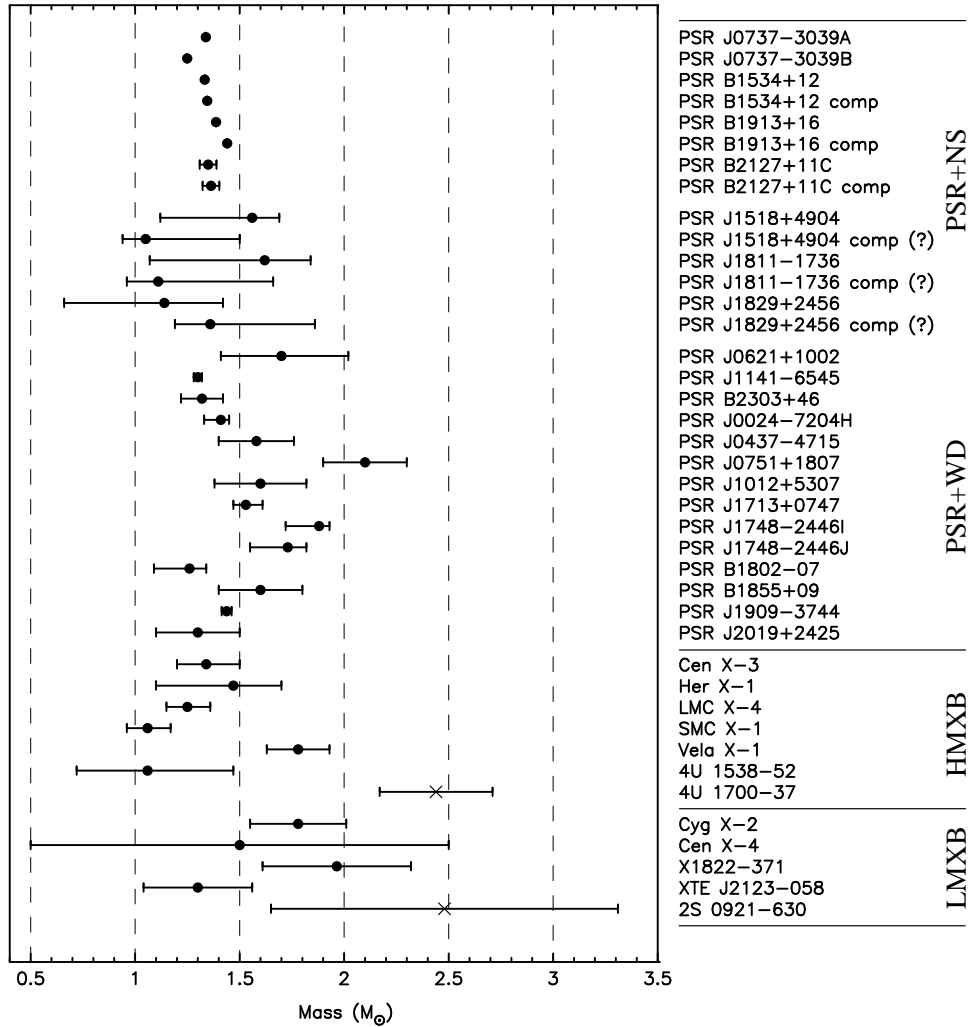,width=13.1cm}   
\end{center}
\caption{Mass measurements of neutron stars in binary systems.
Values for pulsar + neutron star (PSR+NS) and pulsar + white dwarf (PSR+WD) 
systems are from References \cite{TC99} and \cite{S04}, 
with updated measurements of
PSR J0751+1807 from Reference \cite{Netal05},
PSR J1713+0747 from Reference \cite{Setal05},
PSR J1748-2446I \& J from Reference \cite{Retal05}, and
PSR J1909-3744 from Reference \cite{Jetal05}.
Values for the high-mass X-ray binaries (HMXBs) 
Cen X-3, SMC X-1, and LMC X-4 are from Reference \cite{vdMKvKvdH05},
Her X-1 from Reference \cite{vKvPZ95},
Vela X-1 from Reference \cite{Betal01}, and
4U 1538-52 from Reference \cite{vKvPZ95}.
Values for the low-mass X-ray binaries (LMXBs) 
Cyg X-2 is from Reference \cite{OK99},
Cen X-4 from Reference \cite{DAetal05},
X1822-371 from Reference \cite{MDCMP05}, and
XTE J2123-058 from Reference \cite{Setal03}.
The two objects plotted as crosses instead of dots, the HMXB
4U 1700-37 \cite{Clarketal02} and the LMXB 2S 0921-630 \cite{JSNvdK05}
are blackhole candidates, but may be neutron stars.
Error bars are 1$\sigma$ errors. 
}
\label{fig:Masses}
\end{figure}

The masses of almost 40 neutron stars in binary systems have been measured, 
and the results are summarized in  Figure~\ref{fig:Masses}.
The major source of uncertainty usually is the inclination of the orbit with respect to
the line of sight, and it is only in highly relativistic systems that general relativistic
effects allow for a complete characterization of the orbit \cite{TC99}.

The binary systems exhibited in Figure~\ref{fig:Masses} are divided into four classes.
The binary pulsar systems contain a radio pulsar whose companion is either a
neutron star (PSR+NS) or a white dwarf (PSR+WD). 
In the PSR J0737-3039A and B system, both neutron stars are detected as radio
pulsars, whereas in the three systems with the companion marked by ``?''
a white dwarf could be present instead of a second neutron star.
The other two classes are X-ray binaries in which the compact object is accreting
matter from its companion. 
In a high-mass X-ray binary (HMXB) the companion is a massive star, $M_c > 10 M_\odot$,
whereas in a low-mass X-ray binary (LMXB) the companion's mass is below $1 M_\odot$.

\begin{figure}[t]
\begin{center}
\centerline{\psfig{figure=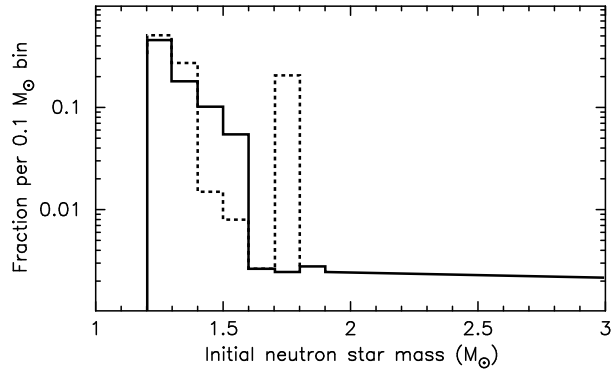,width=8.0cm}}
\end{center}
\caption{The initial mass function of neutron stars as predicted by stellar 
evolution theory.
The continuous line shows results from Reference \cite{FK01} and the dotted line is adapted
from Reference \cite{TWW96}.}
\label{fig:IMF}
\end{figure}

In a HMXB, once the companion ends its life in a core collapse supernova, producing
a second neutron star, the system may be disrupted or remain bound, in which case it
will most probably emerge as a PSR+NS binary system.
A LMXB system will end its life once accretion stops, the companion's mass having been
severely reduced, and the system will emerge as a PSR+WD system.
Given the short lifetime of massive stars, accretion in a HMXB lasts at most a million years,
whereas in a LMXB it can last for billions of years.
Because accretion is limited by the Eddington rate
$\dot{M}_\mathrm{Edd} \approx 10^{-8} M_\odot$ yr$^{-1}$ \cite{ST84}, 
the compact stars in HMXBs, and consequently also in PSR+NS systems, have a mass
very close to their original birth mass, whereas in LMXBs, and consequently in PSR+WD systems,
the neutron star's mass may have increased significantly.
Stellar evolution theory \cite{TWW96} predicts that stars with a main sequence mass below 
$\sim 18$--$19 M_\odot$ will produce, when ending their life in a core collapse supernova, 
proto-neutron stars with masses between $1.2$ to $1.5M_\odot$, whereas more massive ones 
will produce remnants with masses $\sim 1.7$--$1.8M_\odot$, or a black hole.
Fallback during the supernova explosion may alter this \cite{FK01}.
In Figure~\ref{fig:IMF} we plot two expected initial mass functions of neutron stars.
These theoretical theoretical initial mass functions are in good agreement with the 
measurements shown in
Figure~\ref{fig:Masses}, where PSR+NS masses are compatible with all PSR+NS in the
range of $1.2$--$1.5M_\odot$ as well as HMXBs, with the possible exception of Vela X-1,
which may be a representative of the second predicted peak at 1.7--1.8 $M_\odot$.
More massive neutron stars are expected in LMXBs and their PSR+WD 
offsprings, which is confirmed by the very recent high-mass measurement of
PSR J0751+1807 \cite{Netal05}.

\subsection{Temperature and Radius Measurements} \label{sec:obs-cooling}

\subsubsection{Elementary modeling of Neutron Star Thermal Emission}

\begin{figure}[t]
\begin{center}
\centerline{\psfig{figure=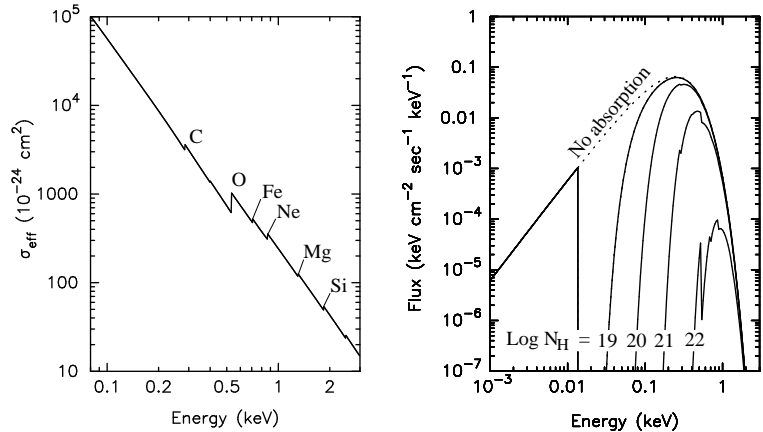,width=10.0cm}   }
\end{center}
\caption{({\em Left}) Effective cross section for interstellar absorption, assuming a
typical interstellar medium chemical composition. Strong edges from various
elements are indicated. Figure taken from Reference \cite{MMcM83}.
({\em Right}) Observable spectrum $f_\infty(E_\infty)$
from a 1.4 $M_\odot$ neutron star with a 10 km radius, $R_\infty$ = 13.6 km, 
at a distance of 500 pcs, assuming blackbody emission at $T_\mathrm{e\, \infty} = 10^6$ K. 
Four cases interstellar absorption with
$N_H = 10^{19}$, $10^{20}$, $10^{21}$, and $10^{22}$ cm$^{-2}$ are shown,
as well as the case of no absorption.
Notice that below 13.6 eV all spectra are identical.}
\label{fig:Sigma-BB}
\end{figure}

The radiation from a point at the surface of the star is specified completely
by the specific intensity $I(\vec{m},E)$, defined such that
$I(\vec{m},E)d\Omega \, dE \, dA \, dt$ is the energy radiated during a time $dt$ 
from an area $dA$ as photons of energy in the range $[E,E+dE]$, which are emerging 
in the direction $\vec{m}$ within a solid angle $d\Omega$ \cite{ST84}.
If the radiation is polarized, e.g., in presence of a magnetic field,
one may separate $I$ into two components corresponding to the two polarization modes.
By integrating over $d\Omega$ in the outward direction, one obtains the specific, 
or spectral, flux
$F(E) = \int\!\!\int d\Omega \, \cos \delta \, I(\vec{m},E)$,
where $\delta$ is the angle between $\vec{m}$ and the outward normal $\vec{n}$ to the 
stellar surface.
Then the flux $F=\int dE \, F(E)$.
The star's luminosity is then obtained by integration of $F$ over the whole stellar surface,
$L = \int\!\!\int dA \, F = 4\pi R^2 F$,
where $F$ must an average value over the stellar surface in case it is not
uniform.
For blackbody emission and a uniform surface temperature $T$, we have $F=\sigma_\mathrm{SB} T^4$ 
and hence $L=4\pi R^2 \sigma_\mathrm{SB} T^4$.
For nonblackbody or/and a non-uniform emission, it is custumary to express
the luminosity as $L=4\pi R^2 \sigma_\mathrm{SB} T_e^4$, a relation that simply 
defines the star's effective temperature $T_e$.

\begin{figure}[t]
\begin{center}
\centerline{\psfig{figure=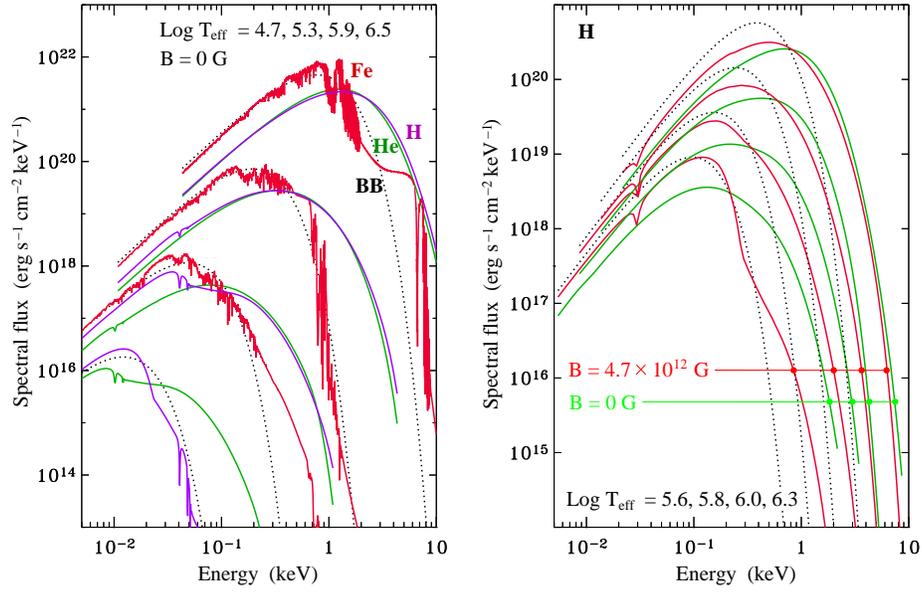,height=8.0cm}}
\end{center}
\caption{({\em Left}) Spectral fluxes of emerging radiation from nonmagnetized atmospheres
of various chemical compositions and effective temperatures.
({\em Right}) Spectral fluxes of emerging radiation from magnetized hydrogen atmospheres
of various effective temperatures and magnetic field perpendicular to the surface.
Figure taken from Reference \cite{ZP02}.}
\label{fig:ZP-Spec}
\end{figure}

Owing to gravitational redshift, a photon emitted at energy $E$ will be detected 
by an observer at infinity with energy $E_\infty = e^\Phi E$ where 
(for a non-rotating star in the Schwarzschild metric)
\be
e^\Phi \equiv g_{00}^{1/2} = \sqrt{1-\frac{2GM}{Rc^2}} < 1 \; .
\label{eq:ephi}
\ee
The luminosity $L_\infty = e^{2\Phi} L$ is redshifted twice because it
has both energy and time content. Similarly, the fluxes at infinity are 
$F_\infty = e^{4\Phi} F$ and $F_\infty(E_\infty) = e^{3\Phi} F(e^{-\Phi}E_\infty)$.
A blackbody at temperature $T$, once red-shifted, would still be seen as a 
blackbody, but one at temperature $T_\infty = e^\Phi T$; hence one also defines
the redshifted effective temperature $T_{e \, \infty} \equiv e^\Phi T_e$ so that
\be
L_\infty = 4 \pi R_\infty^2 \, F_\infty 
         = 4 \pi R_\infty^2 \, \sigma_\mathrm{SB} T_{e \, \infty}^4\,.
\label{eq:lum_inf}
\ee
The radius at infinity has to be $R_\infty = e^{-\Phi} R $ for consistency, but also has the
physical interpretation that an observer able to resolve the angular size of the star would
see it as a disk of radius $R_\infty > R$.
Observers commonly summarize the results of an observation's analysis by giving values of 
$T_{e \, \infty}$ and $R_\infty$.
Note that a realistic atmosphere spectrum of temperature $T_e$ does not redshift
into the spectrum at temperature $T_{e \, \infty}$, so that the redshift is in principle measurable
from spectral fits.

Unfortunately, the luminosity $L_\infty$, or the flux 
$f_\infty = L_\infty/4\pi D^2 = (R_\infty/D)^2 F_\infty$, 
with $D$ as the star's distance, is not directly observable because of interstellar absorption.
The probability for a photon of energy $E$ to be absorbed by ionizing an atom in the interstellar
medium is $\exp[-N_H \sigma_\mathrm{eff}(E)]$,
where $N_H$ is the hydrogen column density and $\sigma_\mathrm{eff}(E)$ is the effective
cross section. 
The quantity that is actually observed, the star's spectrum, is hence the specific 
flux $f_\mathrm{obs}(E_\infty)$, given by
\be
f_\mathrm{obs}(E_\infty) = e^{-N_H \sigma_\mathrm{eff}(E_\infty)} f_\infty(E_\infty) =
e^{-N_H \sigma_\mathrm{eff}(E_\infty)} \left(\frac{R_\infty}{D}\right)^2 e^{3\Phi} F(E_\infty e^{-\Phi})
\label{eq:fit}
\ee
in terms of the specific flux $F(E)$ as emitted by the stellar surface.
An example of $\sigma_\mathrm{eff}(E)$, exhibiting its overall $E^{-3}$ dependence,
is shown in Figure~\ref{fig:Sigma-BB}, as well as the effect of a range of values
of $N_H$ on the observable blackbody spectrum of a neutron star.
Considering that most observed cooling neutron stars have temperature of the order of
10$^6$ K or lower and that a $T=10^6$ K blackbody has its peak at 0.28 keV, one sees that
most photons are absorbed unless the star is quite close, i.e. $N_H \leq 10^{20}$ cm$^{-2}$.

A spectral fit of $f_\mathrm{obs}$ with a blackbody model involves only three parameters,
$T_\infty$, $N_H$, and $R_\infty/D$, because using the blackbody spectrum $F_{BB}$ for $F$ in 
Equation~\ref{eq:fit} yields $e^{3\Phi}F_{BB}(T;E)=F_{BB}(T_\infty;E_\infty)$, 
where $e^\Phi$ drops out.
For blackbody emission $T_\infty$ and $N_H$ essentially determine the shape of the 
observed spectrum and $R_\infty/D$ is a scaling factor.
Having determined $T_\infty$ and $N_H$, if the distance $D$ is known,
one obtains a measurement of $R_\infty$.
A reliable measurement requires, however, that the real spectrum $F(E)$
does not differ too much from that of a blackbody.

\subsubsection{Models of The Neutron Star Surface}
\label{sec:spectrum}

The assumption that the locally emitted spectrum $F(E)$ is a blackbody is very dubious.
Romani \cite{R87} was the first to show that that deviations of $F(E)$ from $F_{BB}(E)$ can be
very large when the neutron star surface is an atmosphere.
The extent of the deviations depends strongly on the chemical composition, which, 
unfortunately, is unknown.
For a light-element composition, H or He, the opacity decreases strongly with photon energy, 
$\sim E^{-3}$, and high-energy photons emerge from deeper, and hence hotter, layers:
The resulting spectrum  has a strong excess compared with a blackbody's Wien tail and a strong
deficit in the Rayleigh-Jeans regime, as can be seen in the examples
shown in the left panel of Figure~\ref{fig:ZP-Spec}.
For an atmosphere composed of heavier elements \cite{ZPS96,NoBSpectra}, 
the opacity energy dependence is not
as strong and the emerging spectrum is closer to a blackbody than in the case of light
elements, but presents numerous absorption lines, as, e.g., in the Fe atmosphere spectra
shown in Figure~\ref{fig:ZP-Spec}.
In the presence of a strong magnetic field, as is the case for all isolated cooling neutron stars,
the specific intensity becomes strongly anisotropic \cite{PSVZ94} and the resulting spectra are,
usually, intermediate between a blackbody and the corresponding nonmagnetized spectra
\cite{RRC97-M92,MagHHE}.
Examples of magnetized $H$ atmosphere spectra are shown in the right panel of
Figure~\ref{fig:ZP-Spec}.

\begin{figure}[t]
\begin{center}
\centerline{\psfig{figure=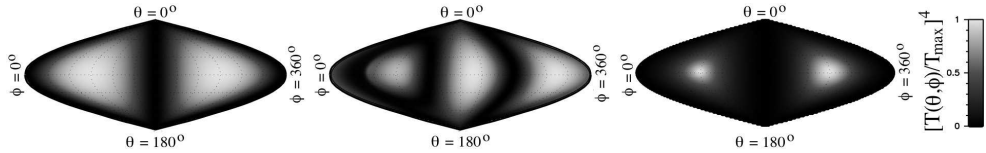,width=13.0cm}   }
\end{center}
\caption{Expected surface temperature distribution for three different magnetic 
field configurations. 
In the left panel the field is a pure poloidal dipole with its north pole at 
$(\theta,\phi) = (90^\circ,90^\circ)$ and a strength of 10$^{12}$ G.
In the central panel a poloidal quadrupolar component has been added to the poloidal 
dipole, whereas in the right panel a todoidal component, inmersed deep in the crust and
with the same symmetry axis as the poloidal component, has been added.
Figure taken from References \cite{PGW06} and \cite{GKP06}.}
\label{fig:Tsurf}
\end{figure}

Given the strong increase in electron binding energies in the presence of a strong magnetic field,
it is also possible that the neutron star surface may be in a condensed state, i.e., 
a liquid or a solid \cite{Lai01}.
Models of the spectrum emitted by such a surface have been presented only recently
\cite{MagSurf}, and for some specific field strength and chemical composition, it may simulate
a blackbody in the observable energy range of 0.1--1 keV.

The presence of a strong magnetic field is not only felt 
at the surface, but also felt much deeper in the star's crust where heat from the interior is
transported to the surface by electron conduction.
Thermal transport transverse to the field is strongly suppressed so that heat essentially
flows along the magnetic field lines, resulting in a nonuniform surface temperature distribution
\cite{GH83,P95}.
When observing the neutron star thermal radiation in the X-ray band, one preferentially detects
photons from the warm regions of the surface.
If extended cold regions are present at the surface, they may nevertheless dominate the thermal
flux at low energies, i.e., in the optical range.
Examples of possible surface temperature distributions owing to the anisotropic thermal 
conductivity are illustrated in Figure~\ref{fig:Tsurf}.

\subsubsection{Temperature and Thermal Luminosity Measurements of Isolated Cooling Neutron Stars}
               \label{sec:cooldata}

\begin{figure}[t]
\begin{center}
\centerline{\psfig{figure=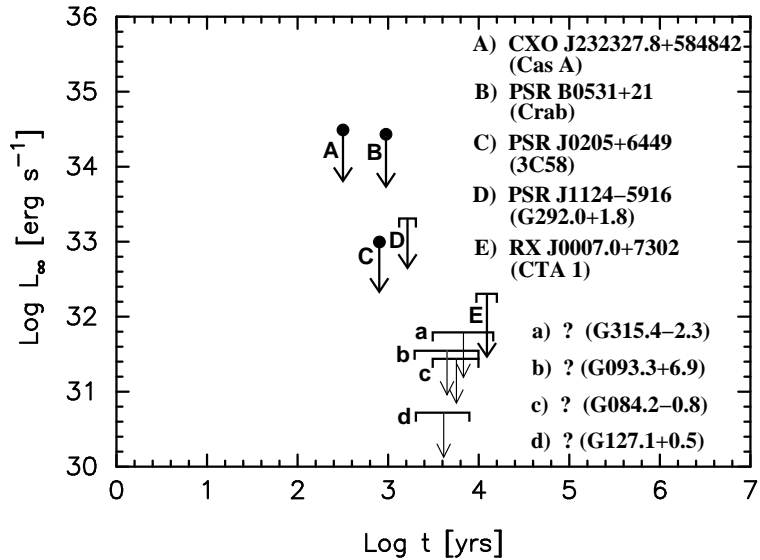,width=10.0cm}}
\end{center}
\caption{Upper limits on the thermal luminosities of neutron stars 
in supernova remants (names are in parenthesis).
Figure adapted from Rerefence \cite{PLPS04}, with Reference \cite{Weisskopf2004}}
\label{fig:Cool_Data_Upper}
\end{figure}

\begin{figure}[t]
\begin{center}
\centerline{\psfig{figure=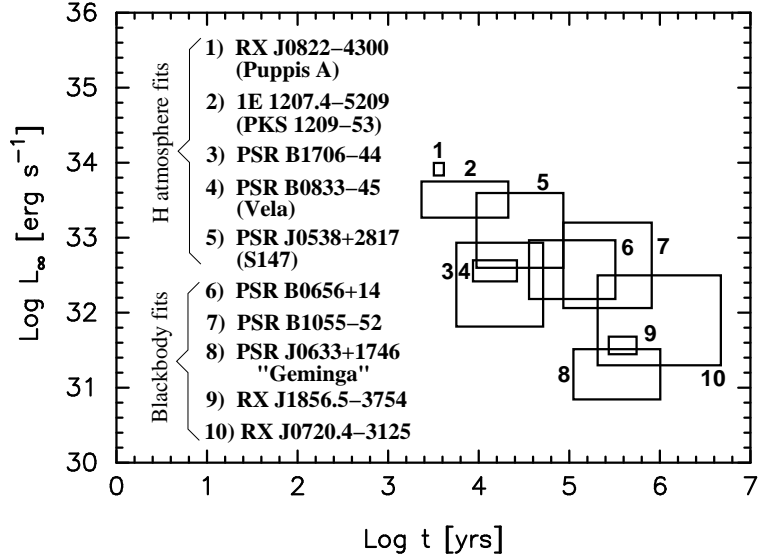,width=10.0cm}}
\end{center}
\caption{Measured thermal luminosities of isolated cooling neutron stars.
The names of the associated supernova remnants, when such a remnant is known,
are given in parentheses. 
Figure adapted from Reference \cite{PLPS04}}
\label{fig:Cool_Data_Good}
\end{figure}

The very brief description presented above gives a flavor of what is involved in
interpreting observational data about cooling neutron stars and what is needed to obtain reliable 
measurements of their temperatures, thermal luminosities, and, we hope, radii.
We summarize in Figures~\ref{fig:Cool_Data_Upper} and \ref{fig:Cool_Data_Good} 
the best presently available results.
All objects shown as PSR are active pulsars and have a very energetic magnetosphere 
-- which is a copious X-ray, and in some cases also $\gamma$-ray, emitter --
producing a power-law spectrum superposed on the thermal emission.
Separating the surface thermal emission from this magnetospheric emission is not a trivial task.
For the Crab pulsar, PSR B0531+21 and PSR J1124-5916, 
the magnetospheric emission is so strong that
the thermal emission is undetectable and only upper limits on the latter are obtainable.
In the similar case of PSR J0205+6449 in 3C58, the thermal emission is barely detected as a slight
correction to a pure power-law spectral fit.
RX J0007.0+7302 in the supernova remnant CTA1 manifests itself as a point source with a power-law
spectrum embedded in a pulsar wind nebula, and no pulsations have been detected to date: 
The thermal emission is also undetected and hence only an upper limit on it is possible.
The four putative objects in Figure~\ref{fig:Cool_Data_Upper} marked ``?''
come from a deep search for central objects in these four supernova remnants \cite{Kaplan2004}:
No compact object has been found and the quoted upper limit is very
restrictive if the compact object is a neutron star, but the latter may also be a black hole.

For the stars in which the thermal spectrum is clearly detected, the dilemma is in the choice of the
theoretical spectrum to be used in the fits.
In the results shown in Figure~\ref{fig:Cool_Data_Good} we distinguish between spectral fits 
performed with blackbody and those performed with magnetized hydrogen atmosphere models,
depending on the resulting $R_\infty$ estimate.
Blackbody fits give higher $T_{e \, \infty}$ and smaller $R_\infty$ than atmosphere model fits:
An atmosphere model fit is chosen when a blackbody fit gives too small an $R_\infty$, and a
blackbody fit is chosen when an atmosphere model fit gives a much too large $R_\infty$.
Moreover, in several of the cases where the atmosphere model fit is prefered, the blackbody fit
is statistically unacceptable.
Considering, moreover, that only the hottest part of the star may be detected in the X-ray band,
the deduced temperature may not be $T_{e \, \infty}$, as defined in Equation~\ref{eq:lum_inf}.
The values of $L_\infty$ we report in Figure~\ref{fig:Cool_Data_Good} take into account this 
ambiguity and are probably more reliable observational values to compare with theoretical cooling
calculations than $T_{e \, \infty}$, a point of view we adopt in Section~\ref{sec:section5}.

The two stars RX J1856.5-3754 and RX J0720.4-3125 are two of the Magnificent Seven
X-ray dim isolated neutron stars, all of which
share the peculiarity that the blackbody fit of their X-ray thermal spectra results in $R_\infty$
much smaller than 10 km \cite{HaberlReview}.
However, optical detections show in several cases a clear Rayleigh-Jeans spectrum 
corresponding, when fitted with a blackbody, to a much lower temperature than the
X-ray-detected spectrum and to a much larger emitting area \cite{HaberlReview,Pons:2002}.
They may be candidates for neutron stars having strong internal toroidal fields,
as illustrated in the right panel of Figure~\ref{fig:Tsurf}, and reinforce the case that,
to date, radii inferred from spectra of strongly magnetized neutron stars are unreliable.
For example, RX J1856.5-3754 was proposed as a candidate for a quark star \cite{QuarkStar},
but the optical data do not support an anomalously small radius 
\cite{Pons:2002,NoQuarkStar}.

\subsubsection{Radius Measurements of Neutron Stars in Quiescent Low-Mass X-ray Binaries}
\label{sec:qLMXB}

It is clear from the previous discussion that measurements of neutron star
radii through fits of their thermal spectra will give reliable results only
if the following three conditions can be met:
(a) The chemical composition of the atmosphere is known,
(b) the magnetic field is small enough, $<10^{10}$ G so as to not affect the spectrum, and 
(c) the star's distance can be accurately measured.
Quiescent LMXB{\bf s} in globular clusters fulfill these conditions \cite{RBBPZ2002}.
Owing to past accretion, H is present and, because of the strong gravity,
sedimentation assures that it will float to the surface, whereas heavier elements will  sink,
resulting in a pure H atmosphere.
As in all LMXBs magnetic fields are negligibly small and
globular clusters, containing $\sim 10^5$--$10^6$ stars, have distances that can be
measured with an accuracy of 5-10\%.
A possible drawback is that, despite the system being quiescent, it is not yet
possible to exclude that accretion is still occurring at a very low rate,
adding constantly to the atmosphere heavy-elements that significantly alter the spectrum.
Being in themselves very interesting systems, many globular clusters have been, 
and are still observed by CHANDRA and XMM-Newton,
which found \cite{Heinke2003} more than 20 quiescent LMXBs (qLMXBs),
and more candidates are constantly being detected.

The radii of the two qLMXBs in globular clusters $\omega$ Centauri and M13, 
which show purely thermal spectra, have been recently estimated as
$R_\infty = 13.6\pm0.3$ km \cite{Gendre2003a} 
and $R_\infty = 12.8\pm0.44$ km \cite{Gendre2003b}, respectively.
However, the fits were realized with a library of H atmosphere spectra from Reference \cite{ZPS96}
which were calculated at fixed surface gravity $g_s=2.43\times10^{14}$ cm s$^{-2}$.
Because $R_\infty$ and $e^\Phi$ are parameters of the spectral fit (Equation~\ref{eq:fit})
and independent variations of them are equivalent to independent variations of $M$ and $R$, 
a self-consistent spectral fit requires a set of model atmosphere spectra $F$ for the whole
range of fitted $g_s$.
Heinke et al. \cite{Rybicki06} recently showed that such a self-consistent analysis
leads to significant changes in the deduced $R_\infty$, and we show in Figure~\ref{fig:mr}
their results for the qLMXB X7 in the globular cluster 46 Tucanae.

\subsection{Transiently Accreting Compact Stars}
\label{sec:SXRT}

The soft X-ray transients (SXRT), also known as X-ray novae \cite{XrayNovae},
are a class of LMXBs in which accretion is not continuous.
They present repetitive phases of high accretion separated by periods of quiescence.
During outbursts they reach very high luminosities, $L_o \sim 10^{37}$--$10^{38}$ erg s$^{-1}$,
whereas in quiescence the luminosity drops by many orders of magnitude, $L_q \leq 10^{33}$ erg s$^{-1}$.
The typical duration of an outburst, $t_o$, is much shorter than the recurrence time
between outbursts, $t_r$.
A common interpretation of this bimodal behavior is the disk instability model that was
developed to explain dwarf novae \cite{Osaki1996}, which are similar systems in which the
accreting star is a white dwarf, and extended to SXRTs \cite{Lasota2001}:
The companion star is losing mass at a very low rate and feeding a disk that
becomes periodically unstable and empties rapidly onto the central star,
producing a spectacular outburst.
Many SXRTs contain black hole candidates, but in some cases Type I X-ray bursts occur and 
we can be certain that the accreting object is a compact star. 
Below we consider only the systems that are known to contain a compact star.

During the quiescent phases between outbursts, many of the SXRTs show a thermal X-ray spectrum
corresponding to surface temperatures of the order of $10^6$~K:
We see the surface of the neutron star that has been heated during 
the accretion phase(s).
Given the high internal temperature of the star, the heat released in the upper layers
by accretion and thermonuclear burning flows back to the surface,
because of the large temperature gradient in the envelope, and is radiated away.
Non-equilibrium processes inside the star, induced by the compression of matter
because of accretion, release heat that is stored in the stellar interior and that 
leaks out slowly when accretion stops.
Such processes certainly occur within the crust:
Iron-peak nuclei produced by thermonuclear burning at the surface, when pushed to higher
densities, undergo a series of reactions
-- electron capture, neutron emission and pycnonuclear fusion \cite{HZ} -- 
until they dissolve into the star's core.
Overall, an amount of heat, $Q_\mathrm{nuc} \approx 1.5$ MeV, is released in the crust
for each accreted baryon.
Brown et al. \cite{BBR98} showed that this energy is more than sufficient to explain the 
observed quiescent (thermal) luminosities $L_\mathrm{q}$  
in terms of the time-averaged accretion rate, taken over many accretion cycles,  $<\!\dot{M}\!>$:
\be
L_\mathrm{q} \simeq  f \; Q_\mathrm{nuc} \frac{<\!\dot{M}\!>}{m_\mathrm{u}} \simeq 
6 \times 10^{32} \; f \;\frac{Q_\mathrm{nuc}}{1.5 \, \mathrm{MeV}} \,
\frac{<\!\dot{M}\!>}{10^{-11} M_\odot \mathrm{yr}^{-1}} \; \mathrm{erg} \, \mathrm{s}^{-1} ,
\label{Eq:f}
\ee
where the coefficient $f$ represents the fraction of $Q_\mathrm{nuc}$ which is actually stored 
in the stellar interior, i.e., not lost by neutrino emission.  
The luminosity $L_\mathrm{o}$ during an accretion outburst can be estimated as
$L_\mathrm{o} \simeq (\Delta M / t_\mathrm{o}) (GM/R)$,
where $\Delta M$ is the mass accreted during the outburst,
and similarly, $<\!\dot{M}\!>$ can be estimated as $\Delta M/t_\mathrm{r}$.
From Eq.~\ref{Eq:f} one obtains
\be
L_q \simeq L_o \times f \frac{t_o}{t_r} \left(\frac{GM}{Rc^2}\right)^{-1} \frac{Q_\mathrm{nuc}}{m_uc^2}
    \simeq L_o \times f \frac{t_o/t_r}{100 \; \mathrm{to} \; 200} \; ,
\label{eq:BBR}
\ee
where the 100 to 200 range corresponds to the possible range of $2GM/Rc^2$
from $\sim0.33$ to $\sim0.66$, or
\be
f \simeq \frac{L_q}{L_o} \cdot \frac{t_r}{(100 \; \mathrm{to} \; 200) \times t_o}
\label{eq:BBR2}
\ee
Note that $L_\mathrm{o}/L_\mathrm{q}$ is independent of the source's
distance, which is often poorly constrained, whereas $t_\mathrm{o}$ and
$t_\mathrm{r}$ can in principle be obtained directly by monitoring the
source for a long enough time.

\begin{table} 
\begin{center}
\caption{Storage efficiency of transiently accreting neutron stars 
         \protect\rule[-2mm]{0mm}{5mm}
         \label{tab:SXRT}}
\bt{lcc}
\hline
\hline
\rule{0mm}{5mm}
Source            & \bt{c} Storage\\
                  efficiency $f$ \et &       Reference(s)       \\
\hline 
\hline 
Aql X-1           &   $\sim  1 $            &  \cite{CSL97}  \\
CX1 (in NGC 6440) &   $\sim 1$              &  \cite{Cackett-etal05} \\
4U 1608-52        &   $\sim 1$              &  \cite{CSL97}     \\
MXB 1659-29 $^a$   &   $\sim 0.1$            &  \cite{Wetal03}    \\
EXO 1745-248      & $\simless 0.01$   &  \cite{Wetal05b}     \\
SAX J1808.4-3658  &  $\sim 0.01$            &  \cite{Cetal02} \\
\hline
Cen X-4           &   $\sim 0.1$      &  \cite{CSL97}     \\
XTE J1709-267 $^a$ & $\simless 0.01$   &  \cite{Jetal04}  \\
XTE J1751-305     & $\simless 0.1$    &  \cite{Wetal05}   \\
XTE J1807-294     & $\simless 0.1$   &  \cite{Campana-etal05} \\
\hline 
XTE J0929-314     & $t_r/$780 yrs           &  \cite{Wetal05,Campana-etal05} \\
KS 1731-260 $^a$   & $t_r/$1,000 yrs         &  \cite{Wetal01}     \\
X 1732-304 $^a$    & $t_r/$500 yrs           &  \cite{Cackett-etal06} \\
EXO 1747-214      & $t_r/$1,500 yrs         &  \cite{Tomsick-etal05}\\
XTE J2123-058     & $t_r/$67 yrs            &  \cite{Tomsick-etal04} \\
\hline 
\et
\end{center}
\noindent 
\footnotesize{$^a$ Quasi-persistent sources: Outbursts last several years instead of only a few
weeks to a few months.}
\end{table}

We present in Table~\ref{tab:SXRT} a list of SXRTs for which enough information is available
to deduce, from Equation~\ref{eq:BBR2}, an estimate of the storage efficiency $f$.
The six systems in the upper part of the Table have been detected
in outburst many times, so estimates of $t_o$ and $t_r$ are reliable, whereas the four cases in the
middle of the table have exhibited only two or three outbursts and for which estimates of $f$
have to be taken with caution.
For the five systems in the lower part of the table, only one outburst has been
detected, $t_o$ is poorly constrained, and $t_r$ is
at best a guess.
Thus we give for $f$ an entry of $t_r/$number-of-years,
i.e., a best estimate of the right side of Equation~\ref{eq:BBR2}.
However, for four of these five systems, an $f$ of order one would require very long 
recurrence times $t_r$, which are well beyond what the disk instability model 
can accommodate \cite{Lasota2001}.
However, the validity of the storage efficiency $f$ rests on the assumption that
the neutron star has reached a quasi-stationary state, which requires fairly regular
recurrent accretion for at least $\sim 10^4$ yrs \cite{CGPP01}, a time scale far
beyond present coverage of the sources !
Notwithstanding these limitations, it is striking that in most cases the storage efficiencies
seem to be quite low and seem to indicate the occurrence of strong neutrino losses.

\subsection{Other Possible Observational Constraints}
\label{sec:Others}

The above selection of observational constraints is of course incomplete and 
reflects our bias in judging relevant and promising probes 
of the physics of the inner core. We very briefly mention here some other possible constraints.

Quasi-periodic oscillations at millisecond periods observed in several LMXBs
may constrain the mass and radius of the neutron star, but interpretation of the origin and
nature of these phenomena is still controversial and inconclusive \cite{vanderKlis:2000}.
Detection of atomic absorption lines in the thermal spectrum of a neutron star can 
directly yield the gravitational redshift due to the star when the line is identifiable.
The identification can be secure only for stars with a magnetic field sufficiently weak so 
that it does not affect the atomic structure.
Candidates for these measurements are the neutron stars in LMXBs, and there is, to date, only one
case of such detection:
Fe XXVI and XXV for n=2--3 and O VIII for n=1--2 transitions in the burst spectrum of 
EXO 0748-676 \cite{Cottam:2002},
which results in a redshift of $z=0.35$, i.e., $e^{-\Phi}= 1.35$
\footnote{Fe XXV and Fe XXVI are Fe with only two and one electrons bound to it, respectively}.
Also, in Type I X-ray bursts from LMXBs, fits of model spectra to the observed burst spectra
can potentially measure both $M$ and $R$ \cite{BurstsReviews}, but model spectra do not seem to
be accurate enough to produce reliable and reproducible results.
The recently discovered superbursts \cite{Superbursts}
are interpreted as the step following the normal short Type I X-ray bursts, 
which are due to explosive burning of $^4$He into $^{12}$C at the surface of an 
accreting neutron star in a LMXB, which consists 
of explosive thermonuclear burning of the accumulated layer of C into iron-peak nuclei.
However, reproducing the observed properties of these superbursts imposes very tight constraints
on the temperature in the crust of these stars and on the neutrino 
emission from the core \cite{Brown:2004}.

In isolated neutron stars, two more constraints emerge.
From the observations of glitches in several young radio pulsars, Link et al \cite{LEL99}
have deduced, under the commonly accepted scenario that these involve differential rotation
of the neutron superfluid in the inner crust, that the relative moment of inertia of 
the crust is at least 1.4\% of the total stellar moment of inertia.
The second constraint comes from the recently discovered pulsar PSR J1748-2446ad \cite{Hessel2006},
the fastest known pulsar to date with a period of 1.39 ms.
For a given assumed mass, it imposes the stringent limit that its equatorial radius
must be smaller than the mass-shedding radius, at which point the centrifugal force
becomes larger than gravity.
Both results translate into $M$ versus $R$ constraints \cite{LEL99,LP:2004}, which we use 
in Section~\ref{sec:Bulk} and Figure~\ref{fig:mr}.

 \section{DENSE NUCLEON MATTER}
           \label{sec:section3}

Over the past decade there have been numerous attempts to compute the bulk properties of nuclear 
and neutron-rich matter. 
These include microscopic many-body calculations using realistic nucleon-nucleon potentials 
and phenomenological relativistic and non-relativistic mean-field theories. 
The former approach employs a potential that provides an accurate description of the measured 
nucleon-nucleon scattering data and uses variational or quantum Monte Carlo
techniques to obtain the EoS (for a review of these methods, see Reference~\cite{Heiselberg:2000dn}, 
and for recent progress in using variational techniques to obtain the EoS of relevance to 
neutron stars, see Reference~\cite{Akmal:1998cf}). 
Variational techniques rely on the parameterization the  nuclear wave-function. 
Correlations induced by interactions between pairs of nucleons are incorporated 
through suitably parameterized pair-correlation operators, which act on the 
symmetrized product of the Fermi gas wave-function. 
In contrast, quantum Monte Carlo treatments such as the Green's function Monte Carlo are not 
limited by the form of the variational wave-function and can potentially include 
all possible correlations in the many-body system. 
This approach is, however, computationally intensive owing to the spin and isospin 
dependence of the nuclear interaction and grows roughly as $2^A A!/(N!Z!)$ 
where A, N, and Z are the baryon number, neutron number, and proton number, respectively. 
Although this method suffers, in principle, from the fermion sign problem, 
Carlson et al.~\cite{CarlsonGFMC} have developed algorithms to minimize its influence 
and have calculated the ground-state energy of 14 neutrons in a periodic box.

The other notable method that employs realistic nucleon-nucleon potentials is the  
Brueckner-Hartree-Fock (BHF) approach. 
Here the bare nucleon-nucleon interaction is used to determine the interaction energy 
for pairs of nucleons. 
The pairs are treated as independent particles, and correlations that arise because of
Fermi statistics are incorporated. 
This method, developed by Brueckner, Bethe, and Goldstone, is nonperturbative in the 
coupling but utilizes a perturbative expansion in the number of independent 
hole lines  to make it tractable \cite{Bethe:1971}. 
Recent studies have shown that the convergence of this expansion can be  
improved if the single-particle dispersion relation is obtained using the 
Hartree-Fock approximation. 
In the lowest-order Brueckner-Hartree-Fock (LOBHF) method, only the lowest-order terms 
in the hole-line expansion are retained. 
The LOBHF results are in fair agreement with those obtained using variational methods \cite{BHF}.  

The realistic nucleon-nucleon potentials are well constrained by a collection of scattering 
data compiled by the Nijmegen group \cite{Nijmegen:1994}. 
However, these potentials are not unique. 
The data probes the scattering for energies below the pion threshold of $350$ MeV. 
Thus, various models for the nucleon-nucleon interaction with different short-distance behavior 
can be constructed to reproduce the elastic scattering phase shifts up to the measured energy. 
These different potentials are equivalent from the point of view of the low-energy data. 
Alternatively, low-energy observables cannot be sensitive to the details of short-distance physics. 
Nonetheless, these potentials can differ in their predictions for the many-body system because 
additional effective three-, four-, five-, etc. body forces can be relevant. 
The strength of these many-body forces is not unique and depends on the assumed form of 
the two-body force at short distances. 
Fortunately, for the realistic two-body potentials employed to date,
the addition of three-body forces is sufficient to describe light nuclei \cite{3Body}. 
Several groups developed an alternate description of nucleon-nucleon interactions 
based on effective field theory (for a review, see Reference~\cite{EFT}). 
This approach allows for a systematic treatment of the short-distance physics by 
organizing the calculation in powers of momenta and identifies a small expansion 
parameter at small momenta. 
It obviates the need to make specific assumptions about the short-distance potential.  
Bogner et al. \cite{Vlowk} have shown that a momentum-space potential with a rather small 
cutoff of $\sim 2 $ fm$^{-1}$ can describe low-energy data. 
Furthermore, they have shown that all realistic models of the nucleon-nucleon potential evolve 
(in the renormalization group sense) to this form. 
Although it is not clear if these low-momentum interactions will make  
many-body calculations more tractable (or less non-perturbative), preliminary work is 
promising \cite{Bogner:2005sn}. 

In the traditional nuclear mean-field models, the relation to nucleon-nucleon scattering is 
abandoned in favor of a phenomenological interaction whose parameters are determined by 
fitting the model predictions (in the mean-field approximation) to empirical properties of 
bulk nuclear matter at nuclear saturation density. 
These mean-field models, such as the non-relativisitic Skyrme model and the relativistic 
Walecka model and its variants, can be viewed as approximate implementations of the 
Kohn-Sham density functional theory, which has been widely studied and used in 
quantum chemistry and condensed matter physics \cite{Furnstahl:2003cd}. 

Empirical properties of bulk nuclear matter are extracted from the analysis of nuclear masses, 
radii, and excitations of large nuclei. 
These analyses reliably disentangle the nuclear bulk properties 
from surface and electromagnetic contributions. 
Without delving into the details and caveats of these analyses,
we simply state the empirical properties for which there is broad consensus:  
(a) Nuclear saturation density (the density at which symmetric nuclear matter 
is bound at pressure $P=0$) is  $\rho_0 =0.15$--$0.16$ nucleons~fm$^{-3}$; 
(b) the binding energy per nucleon at saturation is $16$ MeV; 
(c) the nuclear compression
modulus defined through the relation $K_0=9 dP/d\rho|_{\rho=\rho_0}$ is 
determined to be in the range of $200$--$300$ MeV; 
(d) the nucleon effective mass at saturation density is $0.7-0.8 ~m_N$; and 
(e) the nuclear symmetry energy defined through the relation 
$S=1/2~\partial^2(\epsilon/\rho)/\partial(1-2 x_p)$, where $x_p$ is the proton fraction, 
is in the range of $30$--$35$ MeV. 
This empirical knowledge provides valuable constraints and guidance for models of dense 
nuclear and neutron-rich matter.
  
\subsection{A Simple Model for Nuclear Matter}

To provide a description of nuclei and nuclear matter, Walecka \cite{Walecka:qa}
proposed a field-theoretical model in which the nucleons interact via exchange of 
scalar and vector mesons.
The model has been refined  and used extensively to study nuclear properties in the 
mean-field approximation (for a review, see Reference~\cite{Serot:1984ey}). 
With the intent of providing a pedagogic overview of the  various forces at  play in 
dense nuclear matter, we describe briefly one such model introduced by Boguta \& Bodmer 
\cite{Boguta:xi} (a more elaborate discussion of this model and its application to 
the structure of neutron stars can be found in Reference~\cite{GBOOK}). 

The model proposes that  in dense matter, nucleons interact with effective short-range forces.  
The Lagrangian is given by
\begin{eqnarray}
{\cal L}_N \!=&& \overline{\Psi}_N \! \left( i\gamma^\mu
\partial_\mu-m_N^\ast
-g_{\omega N}\gamma^\mu V_\mu -g_{\rho N}\gamma^\mu
\vec{\tau}_N\cdot \vec{R}_\mu \!\right)\! \Psi_N \nonumber
\\
&&{} +\frac{1}{2}\partial_\mu \sigma
\partial^\mu\sigma-\frac{1}{2}m_\sigma^2\sigma^2-U(\sigma)-\frac{1}{4}
V_{\mu\nu}V^{\mu\nu} \nonumber 
\\ 
&&{} +\frac{1}{2}m_\omega^2V_\mu
V^\mu-\frac{1}{4}\vec{R}_{\mu\nu}
\cdot\vec{R}^{\mu\nu}+\frac{1}{2}m_\rho^2\vec{R}_\mu \cdot
\vec{R}^\mu,
\end{eqnarray}
where $m^\ast_N = m_N-g_{\sigma N}\sigma$ is the nucleon effective mass, which is 
reduced in comparison to the free nucleon mass $m_N$ owing to the scalar field 
 $\sigma$, taken to have $m_\sigma=600$~MeV. 
The vector fields corresponding to the isoscalar omega and {isovector rho mesons are given by
$V_{\mu\nu} = \partial_\mu V_\nu - \partial_\nu V_\mu$, and $ \vec{R}_{\mu\nu}
= \partial_\mu \vec{R}_\nu -\partial_\nu \vec{R}_\mu $ respectively.  
The exchange of these mesons mimics the short-range forces between nucleons. 
In addition to the coupling between nucleons and mesons, a self-interaction between 
scalar mesons given by 
\begin{equation}
U(\sigma)= \frac{b}{3}m_N(g_{\sigma N}\sigma)^3 + \frac{c}{4}
(g_{\sigma N}\sigma)^4\ ,
\end{equation}
where $b$ and $c$ are dimensionless couplings, is introduced to obtain good agreement with 
the empirical value of the nuclear compressibility \cite{Boguta:xi}.  
$\Psi_N$ is the nucleon field operator and $\vec{\tau}_N$ is the nucleon isospin
operator.  
The five coupling constants, $g_{\sigma N}$, $g_{\omega
N}$, $g_{\rho N}$, $b$, and $c$, are chosen, as in Reference~\cite{GBOOK}, to reproduce 
five empirical properties of nuclear matter at the saturation density listed above.

The model is solved in the mean-field approximation. 
Here, only the
time component of the meson fields have nonzero expectation values.
The symbols $\sigma,\omega$ and $\rho$ denote sigma-, omega-, and rho-meson 
expectation values that minimize the free energy given by
\cite{GBOOK}
\begin{eqnarray}
\Omega_{\rm nuclear}(\mu_n,\mu_e)&=& \frac{1}{\pi^2}\left(
\int_0^{k_{Fn}}dk~k^2~[\epsilon_n(k) - \mu_n] + 
\int_0^{k_{Fp}}dk~k^2~[\epsilon_p(k) - \mu_p]  
\right) \,\nonumber \\
&+&\frac{1}{2}\left(m_{\sigma}^2\sigma^2 
- m_{\omega}^2\omega^2-m_{\rho}^2\rho^2\right)+U(\sigma)
- \frac{\mu_e^4}{12\pi^2} \ , 
\label{omeganuc}
\end{eqnarray}
where
\begin{eqnarray}
\epsilon_n(k) &=& 
  \sqrt{k^2+{m^\ast_N}^2} +g_{\omega N} \omega 
  - \half g_{\rho N}\rho \,,\\
\epsilon_p(k) &=& \sqrt{k^2+{m^\ast_N}^2} +g_{\omega N} \omega 
  + \half g_{\rho N}\rho \,,
\end{eqnarray}
are the neutron and proton single-particle energies. 
The single-particle energy at the Fermi surface defines the respective chemical potentials. 
In stellar matter these chemical potentials are related by the condition of weak interaction 
equilibrium ($n+\nu_e\leftrightarrow e^-+p$) given by 
\be
\mu_n=\mu_p+\mu_e
\ee
(note that we have set $\mu_{\nu_e}=0$ since neutrinos are not typically trapped).
Consequently, only $\mu_n$ and $\mu_e$ are independent and correspond to the two conserved charges, 
namely baryon number and electric charge. 
Furthermore, we require that bulk matter be electrically neutral. 
To enforce local charge neutrality, we require the charge density 
$\rho_Q=\partial \Omega_{\rm nuclear}/\partial \mu_e=0$,
which uniquely determines $\mu_e$.  
Therefore, in effect only $\mu_n$ is independent and the free energy or pressure as a function of 
$\mu_n$ completely specifies the EoS of dense matter. 

\begin{figure}
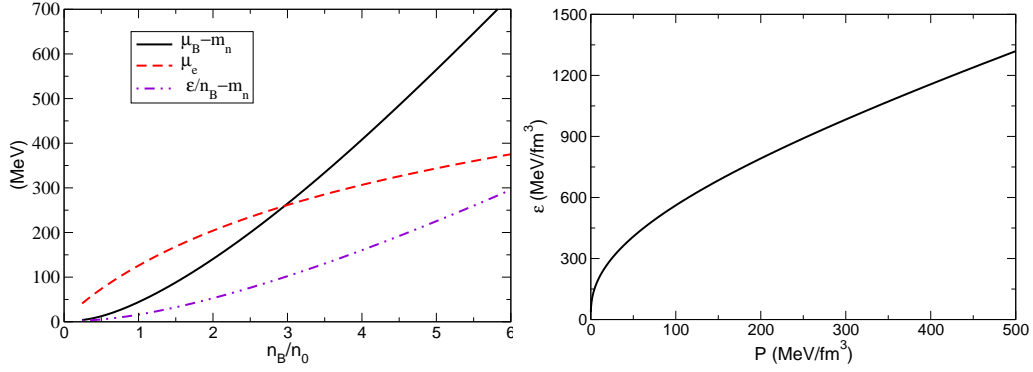

\includegraphics[width=0.5\textwidth,angle=0]{nuceos1.eps}
\includegraphics[width=0.5\textwidth,angle=0]{nuceos2.eps}
\caption{The nuclear equation of state.
({\em Left}) The density
dependence of the baryon chemical potential, the electron chemical
potential, and the energy per baryon. 
({\em Right}) The relation between energy density and pressure.}
\label{nuceos}      
\end{figure}

Figure~\ref{nuceos} shows the thermodynamic properties of charge-neutral stellar matter.  
The left panel shows the baryon chemical
potential, electron chemical potential and the energy per particle as a function of the 
baryon density.
The right panel shows the energy density as a function of the pressure and is usually referred 
to as the equation of state (EoS).  
This relation that is required to
solve for the structure of the neutron star and determines the neutron star mass and radius. 
An EoS that has on average a larger (lower) pressure for a given range of energy density is 
termed stiff (soft). 

Different nuclear EoSs constructed to satisfy the empirical constraints 
at nuclear density and hence similar in this regime can differ significantly 
at lower and higher densities. 
In addition, the difference between symmetric nuclear matter EoS and the neutron-rich 
stellar matter EoS could be significant. 
This difference arises mainly because of the difference in the density dependence of the 
nuclear symmetry energy.  
The magnitude and the density dependence of the proton fraction in particular is sensitive to it. 
In the mean-field model considered here, the nuclear symmetry energy arises owing to the 
isovector force from the exchange of $\rho$ mesons, and its density dependence is linear. 
In contrast, in more sophisticated treatments that employ realistic nucleon-nucleon interactions 
and correlations beyond mean-field theory, the symmetry energy has non-trivial 
density dependence and its magnitude is typically smaller. 

These differences lead to two generic trends that distinguish variational treatments 
such as those reported in the work of Akmal, Pandharipande, and Ravenhall (APR) \cite{Akmal:1998cf} 
from the mean-field EoSs. 
First, the APR EoS for beta-stable matter is considerably softer than the mean-field EoS at 
low density and stiffer at high density. 
Second, the typical proton fraction at high density in the APR EoS is smaller than those 
predicted by the mean-field EoS. 
Both of these trends lead to important consequences for neutron stars. 
The differences in the low- and high-density behavior of the EoS, as we discuss in more detail in 
Section~\ref{sec:section5}, lead to more compact (smaller radii) and more massive (larger maximum mass) 
neutron stars in the APR EoS. 
Because these differences can in part be attributed to differences in the density dependence of 
the symmetry energy, ascertaining this directly from experiment would be desirable. 
Recent work has emphasized that this can be extracted from precise measurements of 
the neutron skin of heavy nuclei \cite{PREX} and heavy-ion collisions \cite{HI}.

The lower proton fraction seen in APR suppresses neutrino cooling of the core. 
This occurs because the single-nucleon neutrino-producing reactions such as 
$n\rightarrow p+ e^-+ \bar{\nu_e}$ and $e^-+p\rightarrow n+\nu_e$, 
which are termed direct URCA reactions, are kinematically forbidden when the proton fraction is small. 
These reactions, which occur on the Fermi surface, can conserve momentum only if 
$|\vec{k_{Fn}}|\le |\vec{k_{Fp}}+\vec{k_{Fe}}|$ (termed the triangle inequality). 
Because electric neutrality requires that  $|\vec{k_{Fe}}|\simeq |\vec{k_{Fp}}|$, it follows that 
the triangle inequality is difficult to satisfy when the proton fraction is small. 
A detailed calculation shows that the direct URCA reaction occurs when the proton fraction is 
$x_p=n_p/(n_n+n_p) \ge 1/9$ \cite{Lattimer:1991}. 
For smaller values of $x_p$, the dominant neutrino-producing reactions involve two nucleons 
in the initial and final state to ensure momentum conservation. 
They include the charged-current processes $nn\rightarrow npe^-\bar{\nu}_e$ and 
$np\rightarrow ppe^-\bar{\nu}_e$ (and their inverse), which are termed modified URCA reactions, 
and the neutral-current processes such as $nn\rightarrow nn \nu \bar{\nu}$, which are 
termed neutrino bremsstrahlung.  
These two nucleon reactions rates are suppressed by a factor of $(T/T_F)^2$ in 
degenerate matter relative to  the direct URCA reaction. 
They are  roughly five orders of magnitude slower for the typical temperatures of interest 
in neutron star cooling (see Section 5).

\subsection{Nucleon Superfluidity and Superconductivity}

The low-energy excitation spectrum and the response properties of the dense nuclear matter 
play key roles in the dynamics and thermal evolution of compact stars. 
As we discuss in Section~\ref{sec:section5}, young neutron stars cool primarily owing
to neutrino emission from the core. 
Thus, phenomena such as superconductivity and superfluidity, which drastically alter the 
low-lying spectrum, become relevant even if their influence on the EoS is not significant. 
Much like electron pairs, via the Bardeen-Cooper-Schreiffer (BCS) \cite{Bardeen:1957mv} 
mechanism in low-temperature solids, nucleons can pair because of their intrinsic attractive 
interactions in nuclei and in dense matter. 
The typical nuclear pairing energy, or gap, is of the order of an MeV. 
Because the ambient temperature inside neutron stars is $\ll $ MeV at all 
times subsequent to a few tens of seconds after their birth, nucleon pairing 
can dominate the low-energy energy response properties of the neutron star interior.     

In the neutron star crust, as nuclei become increasingly neutron rich, neutrons 
drip out of nuclei at a density of $\rho_\mathrm{drip} \simeq 4.3 \times 10^{11}$ g cm$^{-3}$.
At these low densities the neutron-neutron interaction is attractive in the $^1S_0$ channel, 
leading to a large scattering length of $a\simeq 20$ fm. 
This attractive interaction destabilizes the Fermi surface and results in pairing and superfluidity.  
The pairing gap calculated using BCS theory is given by 
\begin{equation}
\Delta = \frac{k_F^2}{M} \exp{\left(\frac{\pi}{N(0)~V(k_F)}\right)},
\label{eq:bcsgap}
\end{equation} 
where $M$ is the neutron mass, $N(0)=M~k_F/\pi$ the density of states at the Fermi surface 
and $V(k_F)$ the interaction potential evaluated at the scale of the Fermi momentum. 
Clearly, the BCS formula provides only an estimate and cannot reliably calculate the magnitude 
of the gap in the strongly interacting system. 
Nonetheless, Equation~\ref{eq:bcsgap} illustrates several key trends seen in more 
sophisticated treatments (for a recent review of superfluidity in dense nucleon matter, see 
Reference~\cite{SFReview}). 
First, at low density where the range of the interaction is unimportant and only the 
scattering length is relevant, we can write $V(k_F) \simeq \pi a/M$. 
Inserting this into Equation \ref{eq:bcsgap} we see that at low density the gap 
increases with increasing density. 
At higher density the range of the interaction becomes relevant, the nucleon-nucleon 
interaction changes character and eventually  becomes repulsive.  
Consequently, the gap will eventually decrease in magnitude and vanish. 
This trend results in a bell-shaped curve for the gap as a function of density. 
Detailed calculations discussed in Reference~\cite{SFReview} indicate that the 
neutron $^1S_0$ gap reaches its maximum value $\simeq 1$ MeV when the neutron Fermi 
momentum $k_{Fn}\simeq 0.8$ MeV, and  vanishes when $k_{Fn} \ge 1.5$ fm.  

Deeper, inside the core, $^1S_0$ proton superconductivity and neutron pairing in 
the ${^3P}_2-{^3F}_2$ channel become possible. 
The typical behavior of the proton superfluid gap and its dependence on the proton 
Fermi momentum is similar to that of $^1S_0$ neutron pairing because the interaction and 
the momentum probed are nearly identical. 
However, because the protons coexist and interact strongly with the dense neutron liquid, 
this changes their 
dispersion relation and consequently the density of states at the Fermi surface. 
In the simple mean-field models, this is encoded in the proton effective mass which 
is significantly reduced. 
Typically, this leads to a factor of 2-3 reduction in the maximum value of $^1S_0$ 
proton gap relative to the $^1S_0$ neutron gap. 
The situation with the ${^3P}_2-{^3F}_2$ neutron pairing in the core is much less understood. 
Even simple BCS estimates that ignore any form of medium-polarization effects, 
such as induced interactions and particle-hole renormalization of the bare potential, 
lead to vastly different predictions for this gap at high density because modern models 
of the nucleon-nucleon potential are not constrained by data at these larger relative momenta. 
This is further exacerbated when medium-polarization effects are taken into account. 
Earlier work indicated that medium-polarization could strongly enhance the gap \cite{large3p2}, 
whereas a recent calculation indicates that a medium-induced spin-orbit interaction 
could lead to a large suppression \cite{small3p2}. 

BCS pairing between nucleons leads to an energy gap  $\Delta$ in the excitation  spectrum. 
Consequently, neutrino processes are exponentially suppressed when $T\ll\Delta$ 
owing to the paucity of thermally excited quasi-particles.  
This would lead us to naively conclude 
that nucleon superfluidity leads to an exponential suppression of the neutrino cooling rates in 
the core, where neutron and proton pairing is likely. 
However, in the vicinity of the critical temperature, the system is characterized by strong 
fluctuations as Cooper pairs form and break. 
These fluctuations give rise to a very efficient neutrino-emission process 
termed the pair breaking and formation, or PBF, 
process \cite{Flowers:1976,Voskresensky:1986}. 
In the BCS approximation for the response function, the PBF process emissivity turns on
just below $T_c$, grows as $T$ decreases, and eventually becomes exponentially 
suppressed when $T\ll T_c$.
For typical values of the $^3P_2$ gap, one finds that at $T \sim 10^9$ K, the PBF rate can
be up to one order of magnitude higher than the modified URCA rate \cite{PLPS04}. 
 
Many-body correlations beyond those induced by the pairing play a role in determining 
the response properties,  especially the weak interaction rates. 
The effects of these correlations on the neutrino emissivity are poorly known. 
For example, until recently the modified URCA reaction and neutrino bremsstrahlung reactions 
were calculated using a simplified one-pion exchange interaction in the Born approximation 
\cite{frim:79}. 
This calculation has been revisited using soft radiation theorems and a more realistic 
and nonperturbative treatment of the nucleon-nucleon interactions\cite{hanh:00}. 
These results indicate a suppression by a factor of 2-3 in the neutrino rates relative to 
the calculation in Reference~\cite{frim:79}. 
The effects due to medium polarization have also been investigated, but the findings have
been inconclusive.  
Researchers suspect that medium-polarization effects through particle-hole correlations
lead to a softening of the pion dispersion relation and are relevant for 
neutrino emissivity  \cite{Migdal:1990vm}.  
In a model calculation, this softening has enhanced the modified URCA rate at 
nuclear density by an order of magnitude \cite{Voskresensky:2001}.  
However, there is still no consensus about the role of these correlations. 
For example, in Ref. \cite{Schwenk:2003pj} the authors find that particle-hole 
screening of the nuclear interaction leads to a suppression in the neutrino bremsstrahlung rate 
\cite{Schwenk:2003pj}. 

From the preceding discussion it should be clear that there are many open issues relating to 
nucleon superfluidity in the core. 
The importance of the PBF process, which operates at relatively small values of the gap 
and dominates neutrino cooling, emphasizes the need to know the precise behavior of the 
$^3P_2$-neutron and $^1S_0$-proton pairing deep inside the core. 
It is also important to revaluate the role of correlations beyond BCS mean-field theory, 
both on the low-temperature specific heat and the neutrino emissivities.

 \section{NOVEL PHASES}
          \label{sec:section4}

With increasing density the chemical potential for baryon number and
negative electric charge increase rapidly owing to the repulsive nature
of strong interactions between nucleons at short distances. This furnishes 
energy for the production of strange baryons and the condensation of mesons.  At higher densities, our knowledge of quantum chromodynamics (QCD) and its asymptotic behavior leads to the expectation that quarks inside nucleons will delocalize and form a uniform Fermi sea of quarks through a deconfinement transition. These expectations are borne from model descriptions. Unfortunately, these model predictions are difficult to quantify and the precise density at which these transitions may occur is poorly understood.  Given the current state of the art, it appears that theoretical advances alone are unlikely to  provide conclusive evidence for a phase transition at supranuclear density. Ultimately, only through their possible presence in compact stars can hope to learn of their existence. This would be possible if we could (a) elucidate in a model-independent manner generic properties of these novel phases and  (b) identify and calculate some properties that would affect observable aspects of the compact star structure and evolution so that they are distinguishable from those predicted for dense neutron-rich matter.  Below we consider a few of these scenarios and discuss key bulk and response properties that distinguish these phases.  

\subsection{Hyperons}

Fig.~\ref{fig:mub} shows the variation of the chemical potential associated with neutral and charged baryons. The thicker curves are predictions of
the mean-field model described in the previous section and the thin
lines correspond to the case wherein strong interactions are ignored. The electrons in dense nuclear matter provide a source for the charge chemical potential. 
Thus the chemical potential for a charged  baryon $B^{\mp}$ is 
$\mu(B^{\mp}) = \mu_B \pm \mu_e$, where $\mu_B \equiv \mu_n$ is the baryonic chemical potential and
$\mu_n$ is the neutron chemical potential. Consequently the source for baryons with different charge differs and negatively (positively) 
charged baryons are favored the most (least).  
The horizontal dashed lines indicate the vacuum masses of the $\Lambda$, $\Sigma^\pm$ hyperons. 
The point at which the corresponding chemical
potentials cross the vacuum masses are also indicated both for the
interacting and non-interacting nuclear cases. 
These points correspond to second-order phase transitions at which specific hyperon species 
begin to be populated. 
If the strong interactions between hyperons and nucleons were attractive (repulsive)
the second-order transition would occur at lower (higher) density. 
It is also clear from the figure that strong interactions between nucleons play an important 
role as it shows that for non-interacting nucleons the transitions occur at 
significantly larger density. 
Hyperon-hyperon interactions can also become relevant especially if they are strongly 
attractive as this could lead to a first-order transition at lower density 
\cite{Schaffner-Bielich:2000yj}.

\begin{figure}[h]
\begin{center}
\includegraphics[width=0.6\textwidth,angle=-90]{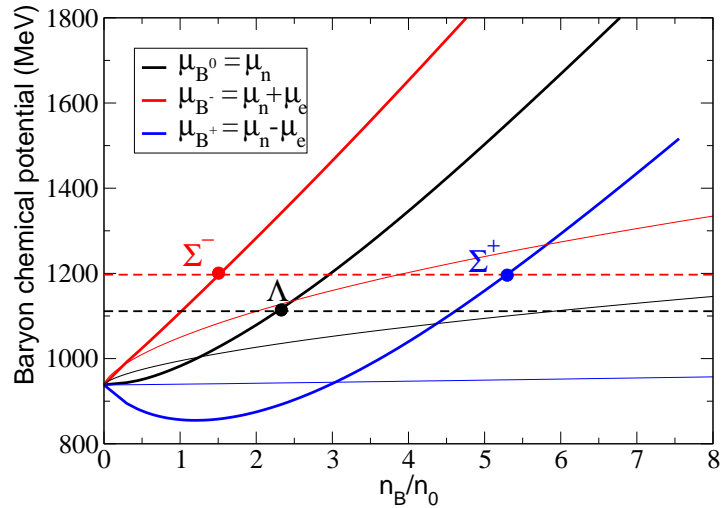}
\caption{Baryon chemical potentials in dense stellar matter.}
\label{fig:mub}   
\end{center}   
\end{figure}

To infer if hyperons exist inside neutron stars it is crucial to either measure or compute directly from QCD the hyperon-nucleon potential.  Experimental data on hyperon-nucleon interactions are very scarce. However, recent developments in studying baryon-baryon interactions using lattice QCD are promising and could potentially provide useful information in constructing a low-energy effective theory
for hyperon-nucleon and hyperon-hyperon interactions \cite{hypQCD}. 
Ref. \cite{Rijken:1998yy} analyzed various experimental constraints on the hyperon-hyperon interaction and Ref.  \cite{Millener:hp} studied constraints arising from the binding energy of the $\Lambda$ in hypernuclei. This information was first employed in the mean-field models by Glendenning \cite{Glendenning:1991es}. Subsequently a more detailed analysis using models for the hyperon-nucleon potentials which are consistent with $\Lambda$ binding energy and the hyperon scattering lengths have been employed in relativistic BHF calculations \cite{Baldo:1998hd}. These studies indicate that the lightest hyperons could appear at relatively low density (2-3 times nuclear density). 

The presence of hyperons typically softens the equation of state and enhances the response of dense matter. This occurs because relative to the nucleons the hyperons contribute more to the energy density than to the pressure because they have larger masses and smaller Fermi momenta. They furnish new degrees of freedom (that are less degenerate than the nucleons) and can readily participate in excitation processes and thereby enhance the response. In particular their presence enhances the neutrino cooling in the core because: (1) they can participate in rapid direct URCA  reactions such as $\Lambda \rightarrow p e^- \bar{\nu}_e$ (not kinematically suppressed) ; and (2) they reduce the neutron to proton ratio and facilitate the momentum conservation needed to initiate the direct URCA involving nucleons \cite{Prakash:1992}.  

\subsection{Kaon Condensation}

A large number of electrons are required to ensure charge neutrality
in dense nuclear matter. The typical electron chemical potential
$\mu_e \sim 100$ MeV at nuclear density. As density increases
$\mu_e$ increases to keep pace with the increasing proton number
density. The magnitude of this increase is sensitive to the
density dependence of the isovector interaction contribution to the
nuclear symmetry energy. In Figure \ref{fig:mue} the density
dependence of the electron chemical potential in the mean-field model
is shown as the thicker curve.  For reference, the
electron chemical potential for the case of noninteracting nucleons
is also shown (thin curve). 
If the energy of a zero momentum negatively charged boson in the medium 
is less than the electron chemical potential it will condense.  
The amplitude of the condensation will in general be regulated by the 
repulsive interactions between bosons.

In the hadronic phase the likely candidates for condensation are 
$\pi^-$ and $K^-$.  In vacuum, pions are significantly lighter
than  kaons but this situation may be reversed in the dense medium
owing to strong interactions between mesons and nucleons. The physical
basis for this expectation is that the effective theory of
meson-nucleon interactions must incorporate the repulsive s-wave
interactions arising owing to the Pauli principle between nucleons and
mesons composed of only up and down quarks, whereas
interactions between mesons with $\bar{u}$ or $\bar{d}$ quarks and
strange quarks must be attractive.
Experiments with kaonic atoms lend strong support to the
aforementioned theoretical expectation \cite{Friedman:hx}. In
Figure \ref{fig:mue} the vacuum pion and kaon masses are shown as the dark and light dashed
lines respectively. 
If the masses do not change in the medium, the figure indicates
that $\pi^-$ condensation occurs in the vicinity of nuclear density
whereas $K^-$ condensation does not occur for the densities
considered. When interactions with medium are included, a uniform
charged pion condensate is disfavored owing to a weak and repulsive
s-wave interaction. Instead, a spatially varying condensate can be
favored because of attractive p-wave interactions (for a review see
Ref. \cite{pionreview}). The kaon-nucleon interaction, however,
is strongly attractive. For this simple reason we discuss kaon condensation below.

\begin{figure}[h]
\begin{center}
\includegraphics[width=0.6\textwidth,angle=-90]{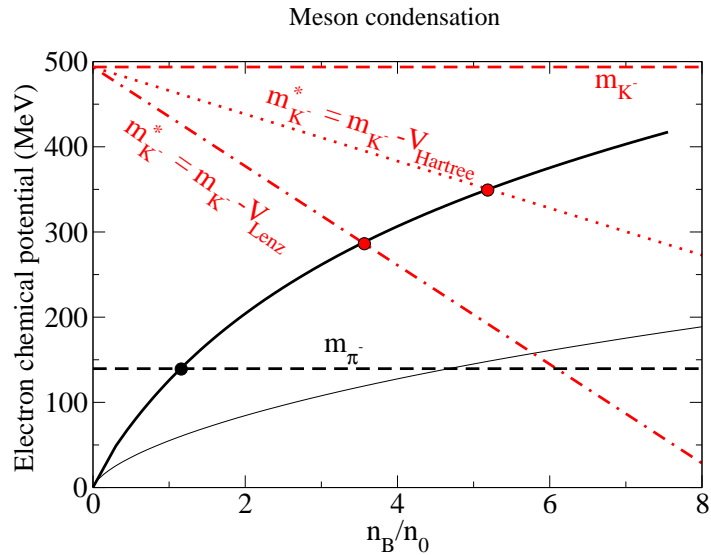}
\caption{Electron chemical potential, the pion and kaon vacuum and
in-medium effective masses in dense stellar matter}
\label{fig:mue}   
\end{center}   
\end{figure}

Kaplan and Nelson \cite{Kaplan:yq} proposed the idea that kaons could condense in dense nuclear matter. Using a simplified SU(3)$\times$
SU(3) chiral Lagrangian they showed that $K^-$ could condense at a
density approximately three times nuclear density. Subsequently, several
authors have studied in detail the nature and the role of kaon
condensation in neutron star matter (for a recent review see
Ref. \cite{Ramos:2000dq}).  Here, we  employ a simple, schematic,
potential model for kaon-nucleon interactions considered in
Ref. \cite{Carlson:1999rr} to illustrate the salient features.  The
scattering length $a_{K^- n}$ characterizes the low-energy
kaon-nucleon interaction and experiment indicates that $a_{K^- n}
\simeq -0.4$ fm. Following Reference \cite{Carlson:1999rr}, 
the interaction potential can be modeled as a square well.  
The parameters of the potential, i.e., the depth $V_0=-122$ MeV and range $R=0.7$ fm, 
are determined by fitting the scattering predictions to low-energy nucleon-kaon data (see Ref. \cite{Carlson:1999rr} 
for further details).  Given this potential the change in the effective mass of $K^-$ 
in a low-density medium of neutrons can be computed using the Lenz approximation. 
In this approximation, the attractive potential energy experienced by a kaon at rest 
can be related related to the scattering length and is given by
\ba
V_{\rm Lenz} = - \frac{2\pi}{\mu}~a_{K^- n}~\rho_n \,,
\label{lenz}
\ea
where $\mu$ is the reduced mass of the neutron-kaon system and
$\rho_n$ is the neutron density. The effective mass of the kaon
computed using Equation~\ref{lenz} is shown in Figure \ref{fig:mue}.  
In this case the kaon effective mass equals the electron chemical when 
$\rho\sim 3-4$ times nuclear density. This corresponds to the critical density
for kaon condensation.  At higher densities, the Hartree or mean-field
approximation should be valid. In this case the attractive
potential energy of the kaon cannot be related directly to on-shell
low-energy kaon-nucleon scattering data. The Hartree potential is given
by
\ba
V_{\rm Hartree} = \frac{4\pi}{3}~ V_0~R^3~\rho_n \,,
\label{hartree}
\ea
where $V_0$ and $R$ are the depth and range of the $nK^-$
potential, respectively.  From Figure \ref{fig:mue} we see that in the
Hartree approximation kaons would condense when $\rho \sim 5-6~\rho_0$ . 

The presence of kaons inside neutron stars influences their structure and evolution 
in a manner similar to that of hyperons. 
The softening of the EoS owing to a kaon condensate is typically larger than the softening by hypernos 
because the zero-momentum kaon condensate contributes to the energy density but not to the pressure. 
By furnishing negative charge it favors a more isospin-symmetric nuclear phase 
containing nearly equal numbers of neutron and protons which results in additional softening. 
As we discuss in \S\ref{sec:section5}, a strong softening in the high-density EoS will result in a 
lower maximum mass and a significantly smaller canonical radius. 
The presence of either pion or kaon condensates result in enhanced 
neutrino cooling \cite{MesonCool}. 
In the case of a kaon condensation enhanced cooling occurs owing to both reactions involving 
kaon decays in the presence of a bystander nucleon and the nucleon direct URCA 
reaction made possible by the larger proton fractions present in the kaon-condensed 
phase \cite{Thorsson:1995}. 

\subsection{Normal and Superconducting Quark Matter}

The occurrence of novel hadronic phases depends on the nature of hadronic interactions and their many-body
descriptions. In contrast, the asymptotic behavior of QCD, which
requires that interactions between quarks become weak with increasing
momenta, strongly supports that at sufficiently
high densities nucleonic degrees of freedom must dissolve to form a
gas of weakly interacting quarks. The precise location of this phase
transition will depend on model descriptions of both the nuclear and
quark phases. This is because all model studies indicate that the
phase transition occurs at rather low densities where perturbative
methods do not apply. The bag model provides a simple description of
the quark matter and the confinement \cite{Bag}. The model was designed to
provide a description of the hadron mass spectrum. The basic tenants
of the model are a non-trivial vacuum and nearly free-quark
propagation in spaces (Bags) wherein the perturbative vacuum has been
restored. This restoration costs energy because it requires the
expulsion of the vacuum condensates. The restoration energy per unit
volume is termed the bag constant and is denoted as $B$. The model
also provides a very simple and intuitive description of bulk quark
matter. The pressure in the bulk quark phase containing up (u), down
(d) and strange (s) quarks is due to the kinetic energy density of
quarks and a negative Bag pressure. At zero temperature this is given by
\ba 
P_{\rm Bag}(\mu_u,\mu_d,\mu_s) = - \sum_{i=u,d,s}
\int_0^{k_{fi}}~\frac{\gamma~d^3k} {(2\pi)^3}\left(\sqrt{k^2+m_i^2} -\mu_i\right)
~ -B \,,
\label{pbag}
\ea
where $\gamma=2_{\rm spins} \times 3_{\rm color}$ is the degeneracy
factor for each flavor,  and $\mu_i$ and $k_{fi}$ are the chemical
potential and Fermi momentum of each flavor, respectively. In the limit of massless
quarks and a common chemical potential for all the quarks, the
pressure has the simple form
\ba 
P_{\rm Bag}(\mu) = \frac{3}{4\pi} \mu^4 - B \,.
\ea
The pressure of bulk quark matter computed using Equation~\ref{pbag}
for $B^{1/4}=150$ MeV and $B^{1/4}=200$ MeV are shown in
Figure \ref{fig:qeos}. The leading-order effects of perturbative
interactions between quarks can also be incorporated in the Bag model
at high density. This has the effect of renormalizing the kinetic
term. At densities of relevance to neutron stars, perturbative
expansion in $\alpha_s=g^2/4\pi$, where $g$ is the QCD coupling constant, 
is not valid. Note that the order
$\alpha_s^2$ calculation of the free energy predicts a behavior similar to that of the bag model
\cite{Freedman:1977gz,Fraga:2001id}.
Recently, Fraga, et al. recomputed the equation of state of massless quark
matter to $O(\alpha_s^2)$. They found the perturbative result is
well reproduced by the following, bag model-inspired, form for
pressure over a wide range of densities relevant to neutron stars
\cite{Fraga:2001id}
\ba
P_\mathrm{perturb}(O(\alpha_s^2),\mu) = \frac{3}{4\pi} a_\mathrm{eff} \mu^4 - B_\mathrm{eff} 
\label{equ:pertquark}
\ea
where $a_\mathrm{eff} = 0.628$ and $B_\mathrm{eff}^{1/4}= 199$ MeV with a specific
choice for the renormalization subtraction point, $\Lambda=2\mu$. 
The pressure obtained using Equation~\ref{equ:pertquark} is shown in Figure \ref{fig:qeos}.  
For reference, the nuclear APR EoS is shown and the crossing points denote the locations of 
the first-order transitions between nuclear matter and quark matter. 
In the extreme case of the bag model with $B^{1/4}=150$ MeV we see that quark matter is 
preferred over nuclear matter everywhere. 
In this scenario, quark matter is the true ground state of matter even at zero pressure 
and is the basis for the strange quark matter (SQM) and strange quark star hypotheses which we discuss below. 

\begin{figure}
\begin{center}
\includegraphics[width=0.8\textwidth]{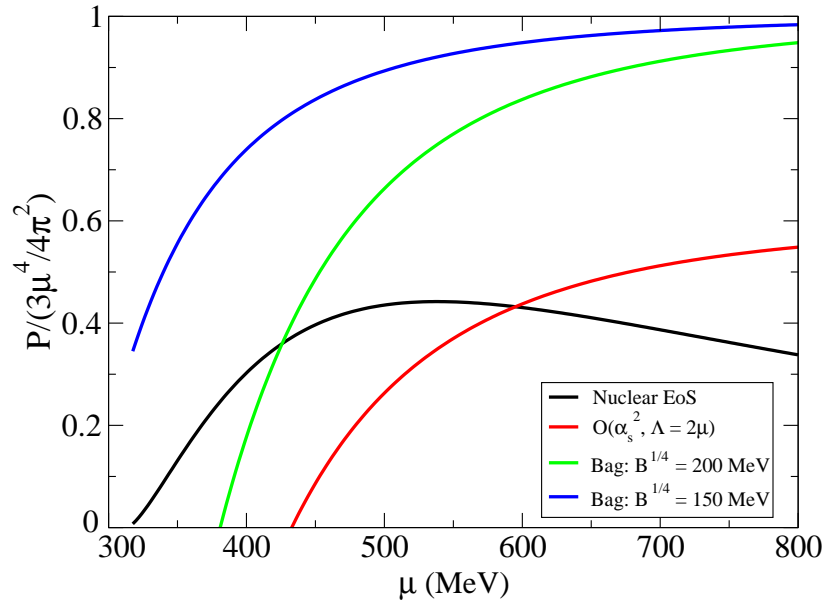}
\end{center}
\caption{Pressure v/s baryon chemical potential for several quark
matter EoS models. The pressure in the nuclear model is also shown for reference.}
\label{fig:qeos}      
\end{figure}

\subsubsection{Color Superconductivity.}

Since the early work of Bardeen, et al., it has been well known 
that degenerate Fermi systems are unstable in the presence of arbitrarily weak 
attractive interactions at the Fermi surface \cite{Bardeen:1957mv}. 
The instability is resolved by the formation of a Bose condensate of Cooper pairs. For the case of charged
fermions, such as electrons, this leads to superconductivity. By analogy,
the presence of attractive interactions between quarks will lead to
pairing and color superconductivity. Barrios and Frautschi \cite{Barrois:1977xd} realized this several decades ago.  The recent realization that the typical superconducting gaps in quark
matter are larger than those predicted in these earlier works has generated
renewed interest. Model estimates of the gap at
densities of relevance to neutron stars suggest that $\Delta \sim 100$
MeV when $\mu=400$ MeV \cite{BigGap} . Here, we provide a 
brief introduction to the subject and emphasize aspects that will 
impact neutron star phenomenology (see Reference \cite{CFLReview} for a more comprehensive recent review).  

One-gluon exchange between quarks is attractive in the color
antisymmetric (triplet) channel and repulsive in the color symmetric
(sextet) channel. The
attraction in the triplet channel can result in s-wave pairing between
quarks in spin-zero and spin-one channels. Explicit calculations show
that the pairing energy is significantly large for the spin zero
case. This type of pairing occurs only between dissimilar flavors of
quarks to ensure that the pair (which is a boson) wave-function is
symmetric. For three massless flavors the condensation pattern that
minimizes the free energy is known as the color-flavor-locked scheme
\cite{Alford:1998mk}. The non-zero condensates in this phase are given by
\begin{eqnarray}
\langle \psi^{i \alpha}_{\rm a,L} \psi^{j \beta}_{\rm b,L}
\epsilon_{ij}\rangle =- \langle \psi^{i \alpha}_{\rm a,R} \psi^{j
\beta}_{\rm b,R}\epsilon_{ij} \rangle &=&
\Delta~\epsilon^{\alpha,\beta, A}~\epsilon_{a,b,A} 
\end{eqnarray}
where $\alpha$ and $\beta$ are color indices, $a$ and $b$ are flavor indices, and
$i$ and $j$ are spinor indices.  
Because $\epsilon^{\alpha,\beta, A}~\epsilon_{a,b,A} =
\delta^{\alpha}_a\delta^{\beta}_b- \delta^{\alpha}_b\delta^{\beta}_a$ 
it follows that condensation locks color and flavor indices. 

In the color-flavor-locked phase all nine quarks participate in
pairing. Consequently there is an energy gap $\sim 2 \Delta$ in the
fermion excitation spectrum. The gluons acquire a mass via the Higgs
mechanism by coupling to the colored condensates. The lowest
excitations in this phase correspond to the Goldstone bosons. The
number and spectrum of Goldstone bosons can be understood by
noting that the condensate breaks baryon number and chiral symmetries. 
Chiral symmetry is broken by a novel mechanism \cite{Alford:1998mk} because independent Left-handed (L) and right-handed (R) condensates  are coupled by the vector interaction.   The octet of
flavor Goldstone bosons that result from the breaking od chiral symmetry acquire a mass because quark
masses explicitly break chiral symmetry. The quantum numbers of these
pseudo-Goldstone modes map onto the meson octet in vacuum. For this
reason they are commonly referred to as the pions and kaons (or
mesons) of the high density phase. Their masses have been computed in
earlier work by matching the effective theory to perturbative QCD at
high-density \cite{CFLmesons}. Unlike in vacuum, the square of meson masses was
proportional to the product of quark masses . This
resulted in an inverted hierarchy, in which the pions were heavier than
the kaons. Explicitly, at high density the masses are given by
\begin{eqnarray}
m^2_{\pi^\pm} &=& a (m_u + m_d)m_s \cr
m^2_{K^\pm} &=& a (m_u + m_s)m_d \cr 
m^2_{K^0} &=& a (m_d + m_s)m_u \,, 
\label{masses} 
\end{eqnarray}
where $a=3\Delta/\pi^2\mu^2$.  It is only the baryon number Goldstone
boson that remains massless.  This mode is responsible for the
superfluid nature of this phase \cite{Alford:1998mk}.

Typically the pairing contribution to the pressure is small, being
$\sim \Delta^2 \mu^2$ because only particles in a shell of width $\Delta$ at the Fermi surface  pair. 
We can supplement the bag model pressure with the pairing contribution due to superconductivity to obtain the pressure of the  CFL phase \cite{Alford:2001zr}
\ba
P_{\rm CFL} = \frac{3}{4\pi^2}~a_{\rm eff} \mu^4 - \frac{3}{4\pi^2}~m_s^2\mu^2 + 
\frac{3}{\pi^2}~\Delta^2 \mu^2 - B\,.
\label{eq:qeos}
\ea
From this equation we can infer that, although the pairing contribution is parametrically small 
by the factor $\Delta^2/\mu^2$ compared to the kinetic pressure, it could become the dominant 
contribution in the region where kinetic pressure and the bag pressure nearly cancel each other. 
Furthermore, because the quark-hadron phase transition is expected to occur in this regime, pairing will likely lower the quark-hadron phase transition density \cite{Lugones:2002,AlfordReddy:2003}. Similar trends are seen in other model calculations such as the Nambu-Jona-Lasino model of 
the quark EoS \cite{NJL}. 

Using Equation~\ref{eq:qeos} with $a_\mathrm{eff}\simeq 0.6$ to describe the quark EoS and the APR 
model to describe the nuclear EoS,  Alford et al. \cite{mimic} find that the quark EoS and 
the nuclear EoS yield nearly the same pressure for a range of energy densities in the vicinity 
of nuclear density. 
This surprising finding, albeit within a specific model, indicates that a phase transition 
at these densities is unlikely to result in significant softening. Nonetheless it would have dramatic consequences for the transport and cooling properties because the 
excitation spectrum of the CFL phase is very distinct from that of nuclear matter.  

\subsubsection{Role of the Strange Quark Mass}

It is a reasonable approximation to neglect the up and down quark
masses at densities of relevance to neutron stars where $\mu \sim 400$
MeV.  The strange quark mass $m_s \sim 200$ MeV, however,
cannot be neglected. When the strange quark mass is large, comparable
to the baryon chemical potential, the Fermi surfaces of the strange
quarks and light quarks will be different. If this difference is
similar to $\Delta$, pairing involving strange quarks is suppressed.
In the limit of infinite strange quark mass, i. e., in the absence of strange quarks, only the light quarks pair. This phase two-flavor superconducting phase is termed 2SC and is also characterized by pairs that are antisymmetric in flavor \cite{BigGap}.  Antisymmetry in color space excludes one of the three colors from participating in the condensation. Thus $SU(3)_{\rm color}$ is broken down to $SU(2)_{\rm color}$ and quarks of a particular color and three gluons remain massless. Furthermore, unlike in the CFL phase no global symmetries are broken. 

Early attempts to bridge the gap between the 2SC and the CFL
phases can be found in References \cite{Alford:1999pa}and \cite{Buballa:2001gj}. The authors found that the CFL pairing scheme was preserved when $m_s \simless \sqrt{2 \mu \Delta}$, whereas for larger $m_s$, a first-order transition to the 2SC phase occurs.  Bedaque \& Schafer showed that the strange quark mass appeared in the effective theory for Goldstone bosons in
the form of a chemical potential for antistrangeness \cite{Bedaque:2001je}. 
When the chemical potential,
$\mu_{\bar{s}} \sim m_s^2/2\mu$, exceeds the mass 
 of $K^0$ which is lightest meson with antistrangeness, these bosons condense in
the ground state.  In the CFL$K^0$ phase hypercharge or
strangeness symmetry is spontaneously broken resulting in the
appearance of a massless Goldstone boson.  For a detailed
discussion of these novel meson-condensed phases and their role in the birth and evolution of neutron stars see Reference \cite{Kaplan:2001qk}.

A finite $m_s$ which is larger than the up and down quark masses leads to a suppression in their number and requires electrons to ensure neutrality. The electron chemical potential needed to accomplish this is $\mu_e \simeq m_s^2/4 \mu$.  This in turn results in a  splitting between the up and down Fermi levels because in beta-equilibrium $\mu_u=\mu-2\mu_e/3$ and $\mu_d=\mu+\mu_e/3$.  In two-flavor quark matter charge neutrality requires a relatively large $\mu_e\sim\mu$ because strange quarks are either absent or too few in number to furnish the necessary negative charge. Consequently, pairing will be favored only in strong coupling ($\Delta \simeq \mu$) and in general this disfavors the 2SC phase in the neutron star context \cite{Alford:2002kj}. However, in  Nambu-Jona-Lasino model calculations of quark matter one typically finds that the effective strange quark mass is large and the quark pairing strength is strong. Both of these effects favor the possibility of realizing the 2SC phase at intermediate densities of relevance to compact stars \cite{NJLStrong}.  If, however, $\Delta \ll \mu$ and pairing between up and down is not possible, then quarks of the same flavor can pair in the spin-one channel. The spin-one phase is characterized by a locking between spin and color, and model predictions indicate a small gap of the order of one MeV \cite{CSL}.

BCS pairing between two species of fermions results in a locking of the Fermi surfaces in momentum space and hence enforces equality in the numbers. This rigidity arises owing to the finite gap which requires a finite energy to produce a quasi-particle excitation. For example, in the CFL phase this locking ensures equal number densities for all three quark flavors even though $m_s > m_{\rm light}$ \cite{Rajagopal:2000ff,Steiner:2002gx}. In the 2SC phase, the number densities of the up and down quarks are equal even when $\mu_e \ne 0$. This persists until $\mu_e \simeq 2 \Delta$ when it becomes possible to generate a number asymmetry at the expense of pairing. However, in simple systems, without long-range interactions, as $\mu_e$ increases one encounters a first-order phase transition from the BCS phase to the normal phase when $\mu_e\ge \sqrt{2} \Delta$ . Surprisingly, even in weak coupling the phase structure of asymmetric fermion systems is not well known.  Several candidate ground states have been suggested. They include(a) the mixed phase, where normal and superconducting phases coexist \cite{Bedaque} (b) the LOFF phase (named after the authors Larkin, Ovchinnikov, Fulde, and Ferrell who first suggested this possibility \cite{LOFF}), where the gap acquires a spatial variation on a length scale $\sim 1/\Delta$ and (c) the gapless superfluid phase (also known as the breached-pair phase) which is a homogeneous superfluid phase with gapless fermionic excitations due to a finite occupation of quasi-particle levels \cite{Liu:2002}. Recent work has shown that in systems with short-range forces the gapless superfluid phase is unstable with respect to phase separation \cite{Bedaque}. However, as we discuss below, all of these possibilities could be relevant in dense quark matter. 

The presence of long-range forces and resulting constraint of charge neutrality in the bulk influences the competition between the various proposed phases of asymmetric Fermi systems. For example although the mixed phase is favored over the gapless phase in neutral systems the neutrality constraint can stabilize the gapless phase with respect to phase separation in two-flavor \cite{gapless2SC} and three-flavor \cite{gaplessCFL} quark matter. These phases which are termed the gapless 2SC and gapless 
CFL phases are remarkable because they are superconducting and yet have a large number of gapless excitations near the Fermi surface. In strong coupling, quantum Monte Carlo results show that gapless phases are stable with respect to phase separation \cite{Carlson:2005}. However, in weak coupling there remain several questions that need to be addressed to ascertain if these phases or minor variants of them are indeed stable. In particular in QCD the magnetic gluon masses are imaginary in the weak coupling gapless phases. This signals an instability known as the chromomagnetic instability and occurs both in gapless 2SC \cite{Huang:2004bg}  and gapless CFL phase \cite{Casalbuoni:2004tb}. Because the Meissner masses can be related to the gradient energy associated with spatial deformation of the pairing field some type of heterogeneity is likely  and it appears that either the mixed phase or the LOFF phase is favored over gapless 2SC \cite{nogapless2SC}. This may also be the case for three-flavor quark matter \cite{Casalbuoni:2005zp}. 

\subsubsection{Absolutely Stable Strange Quark Matter} 

The conjecture that matter containing strange quarks could be absolutely stable is several decades old~\cite{SQM}. This could occur if SQM has a lower energy per baryon at zero pressure relative to heavy nuclei such as $^{56}$Fe, for which the energy per baryon is $\simeq 930 $ MeV. 
If this were true, nuclear matter and nuclei would only be metastable, albeit with a very large lifetime because it would take several simultaneous weak interactions to generate the strangeness needed to lower the energy. However, this situation may not be difficult to achieve in the hot and dense interior of a neutron star. Once strangeness is seeded somewhere in the star it should be able to catalyze the conversion of the entire star on a short time scale, producing a "strange star", i.e., a compact star made entirely of SQM.
For a recent review of SQM and its role in compact stars see Reference \cite{SQMReview}. 

We briefly mention a few recent findings of interest. For SQM to be preferred  over nuclei or nuclear matter its pressure must be larger than the pressure in nuclear matter when the quark chemical potential $\mu \simeq 930/3$. From Equation~\ref{eq:qeos} we see that in the Bag model this would require a relatively small bag constant. However, the bag constant cannot be too small otherwise even two-flavor quark matter would have alower energy than nuclei and these would instantly convert to quark matter as it would not require the simultaneous weak interactions needed to produce strange quarks. In a simple model where $a_\mathrm{eff} \simeq 1, \Delta=0$ and $m_s=150$ MeV a bag constant given by $B^{1/4}\simeq145$ MeV predicts SQM to be absolutely stable while ensuring that the 2 flavor quark matter is unstable. A large gap of the order of $100$ MeV has the effect of increasing the effective bag constant at which quark matter can become stable.  Recent work on the effects pairing in SQM can be found in Ref. \cite{Lugones:2002}. 

Early work indicated that SQM would be homogeneous liquid at arbitrarily low pressure. This is unlike low-density nucleon matter made of nuclei and electrons which is a heterogeneous solid with nuclei embedded in a background electron gas. Reference  \cite{StrangeCrust} investigated the possibility that SQM could also be a heterogeneous solid at low pressure. Using a model-independent analysis this work shows that at relatively low density SQM could also become heterogeneous and phase-separate into positively charged nuggets of quark matter embedded in a background electron gas - much like matter with nuclei at low pressure. With increasing pressure the quark phase would be  populated with voids of electrons.  If stable, the heterogeneous crust could have important consequences for surfaces of strange quark stars. Unlike the conventional scenario for  strange stars which are characterized by 
an enormous density gradient and a large electric field \cite{Alcock} the heterogeneous solid crust would occupy a large radial extent ($\Delta R \sim 50-100$ m) and obviate the need for an electric field.
 
\subsubsection{Excitation spectrum and response properties}
 
Response properties such as neutrino emissivities, viscosity and the low-temperature specific heat have been calculated in both normal and superconducting phases of quark matter. In the normal phase, a large number of particle-hole excitations exist at the Fermi surfaces of the quarks and result in enhanced response and a large specific heat. Consequently neutrino cooling due to processes analogous to the direct URCA reaction but involving quarks is rapid \cite{Iwamoto}. The bulk viscosity is also enhanced owing to the weak interaction rate that converts $d\leftrightarrow s$ quarks \cite{qvisc}.  In contrast, in the CFL phase where all quarks are gapped, the low-lying excitation spectrum is sparse and only the baryon number Goldstone mode contributes at temperatures of relevance to neutron star cooling. Consequently the neutrino emissivity \cite{CFLnuemis}, specific heat and bulk viscosity \cite{CFLVisc} are all relatively negligible. This suppression in the neutrino processes also persists in the  CFL$K^0$ phase\cite{CFLK0nuemis}.  In less-symmetric superconducting phases such as 2SC where ungapped quark excitations exist the situation is similar to that in the normal quark matter. 

The existence of either gapless color superconductivity or the  heterogeneous phases such as the LOFF phase or the mixed phase 
can lead to potentially observable consequences.  In the gapless phase the number of gapless excitations  is anomalously large. The resulting specific heat is a factor $\sqrt{\Delta/T}$ larger than even the normal phase and could potentially impact neutron star cooling at late time \cite{Alford:2004zr}. The coexistence of heterogeneity  and superfluidity is key to understanding glitches.
In this sense  the heterogeneous superconducting phases could play a role. These connections, between the properties of the phase and the observables, are still in their infancy and warrant further work. 

\subsection{Nature of the Phase Transition} \label{sec:pasta}

The novel phases discussed above could occur either via a first- or
second-order phase transition. In first-order transitions phase coexistence is possible. A heterogeneous phase with coexisting phases is termed the mixed phase.  Here charge neutrality is enforced globally. Each of the two coexisting phase have opposite electric charge and the volume fractions adjust to ensure overall neutrality. 

Heterogeneous phases are commonplace in the terrestrial context. 
For example a solid can be viewed as a mixed phase composed of positively nuclear matter  
(residing inside nuclei) coexisting with a negatively charged electron gas phase. 
In the relatively low-density region of the neutron star crust we encounter a 
similar mixed phase that upon increasing density changes to 
accommodate neutrons in the electron gas phase. 
The physics of the neutron-rich mixed phase in the crust is reviewed in detail 
in Ref. \cite{Pethick:1995di}. 
Glendenning noted that similar considerations apply to first-order transitions in 
the context of high-density phase transitions. 
This applies in general to all first-order phase transitions with two 
conserved charges \cite{Glendenning:1992vb}. 
In the neutron star context the conserved quantities are baryon number and electric charge. 
Unlike simple first-order transitions such the water-vapor transition which is 
characterized by one conserved charge (number of water molecules) 
where phase coexistence occurs at specific value of the pressure,  
the phase coexistence in dense matter occurs over a finite interval 
in pressure owing to the presence of the extra degree of freedom, namely electric charge. 

To illustrate the physics of first-order phase transitions and the
role of surface and Coulomb energies in the mixed phase we consider an
explicit example. The phase transition from nuclear matter to CFL
quark matter is a first-order phase transition. The nuclear phase has
no strangeness and the bulk quark phase has no electrons. The
possibility of phase coexistence between these phases was investigated
in Ref. \cite{Alford:2001zr}. We highlight some of the main findings here. In
Figure~\ref{bigpicture}, the pressure of the bulk nuclear, bulk CFL and mixed
phases are shown as a function of $\mu$, the quark chemical
potential. At intermediate values of $\mu$, the mixed phase has larger
pressure and is therefore favored over both the nuclear and CFL bulk
phases. The electron chemical potential, $\mu_e$, required to ensure
charge neutrality in the bulk nuclear phase, which grows with $\mu$,
is shown. In the mixed phase neutrality requires a positively charged
nuclear phase and a negatively charged CFL phase. This is easily
accomplished by lowering $\mu_e$. The decreasing $\mu_e$ in the mixed
phase shown in the figure is obtained by requiring equal pressures in
 both phases at a given $\mu$. 

\begin{figure}[t]
\begin{center}
\includegraphics[width=0.8\textwidth]{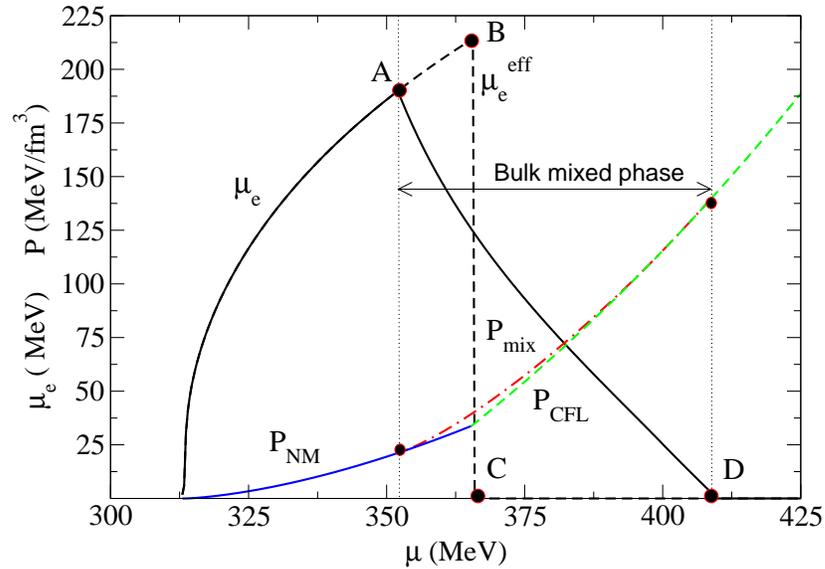}
\end{center}
\caption{Behavior of the electron chemical potential and the pressure
of homogeneous neutral nuclear and CFL matter and of the mixed phase,
all as a function of the quark chemical potential $\mu$. }
\label{bigpicture}
\end{figure}

In the mixed phase the Coulomb and surface energy
costs must be met. The results shown in Figure~\ref{bigpicture} ignored these
corrections. Only bulk
free energy is included; surface and Coulomb energy is neglected.  The
mixed phase occurs between A and D.  The vertical line connecting
B and C denotes the $\mu$ at which the pressures of neutral CFL
and nuclear matter are equal.  This is where a sharp interface may
occur.  The pressure of the mixed phase exceeds that of neutral CFL or
neutral nuclear matter between A and D.  Were this the whole
story, the mixed phase would evidently be favored over the sharp
interface.

In a simple description of the mixed phase one considers
a thin boundary between the coexisting phases. A unit cell of the
mixed phase is defined as the minimum size region that is electrically
neutral. Three different simple geometries are considered: spheres
($d=3$), rods ($d=2$) and slabs ($d=1$) \cite{pasta}. In each of these
cases, the surface and Coulomb energy cost are given by
\begin{eqnarray}
E^S &=& \frac{d~x~\sigma_{\rm QCD}}{r_0} \,, \label{surface}\\
E^C &=& 2\pi~\alpha_{\rm em}f_d(x)~(\Delta Q)^2~r_0^2 \,, 
\label{coulomb}
\end{eqnarray}
where $d$ is the dimensionality of the structure ($d=1,2,$ and $3$
correspond to Wigner-Seitz cells describing slab, rod and droplet
configurations, respectively), $\sigma$ is the surface tension and
$\Delta Q=Q_{\rm nuclear}-Q_{\rm CFL+kaons}$ is the charge-density
contrast between the two phases. The other factors appearing in
Equations~(\ref{surface}) and (\ref{coulomb}) are $x$, the fraction of the
rarer phase that is equal to $\chi$ where $\chi \leqslant 0.5$ and
$1-\chi $ where $0.5 \leqslant \chi \leqslant 1$; $r_0$, the radius
of the rarer phase (radius of drops or rods and half-thickness of
slabs); and $f_d(x)$, the geometrical factor of order one that arises in the
calculation of the Coulomb energy \cite{Pethick:1995di}.  The first step in the calculation is to evaluate 
$r_0$ by minimizing the sum of $E^C$ and $E^S$. The result is
\begin{equation}
r_0 = \left[\frac{d~x~\sigma_{\rm QCD}}{4\pi~\alpha_{\rm
em}f_d(x)~(\Delta Q)^2}\right]^{1/3} \,.
\label{radius}
\end{equation}
Using this value of $r_0$ in Equations~(\ref{surface}),(\ref{coulomb})
the surface and Coulomb energy cost per unit volume is obtained
\begin{equation}
E^S+E^C = \frac{3}{2} \left(4\pi~\alpha_\mathrm{em}~d^2~f_d(x)~x^2\right)^{1/3}
~(\Delta Q)^{2/3}~\sigma_{\rm QCD}^{2/3} \,.
\label{sandccost}
\end{equation}
%

\begin{figure}[t]
\begin{center}
\includegraphics[width=0.8\textwidth]{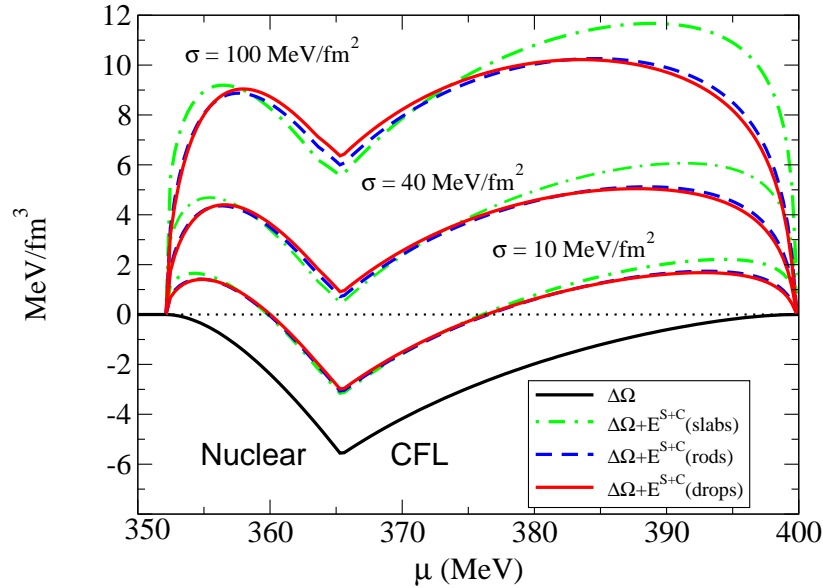}
\end{center}
\caption{The free energy difference between the mixed phase and the 
homogeneous neutral nuclear and CFL phases. In the lowest curve,
the surface and Coulomb energy costs of the mixed phase
are neglected, and the mixed phase therefore 
has the lower free energy. Other curves include surface
and Coulomb energy for different values of $\sigma_{\rm QCD}$ and 
different mixed phase geometry.} 
\label{deltaomega}
\end{figure}

The lowest curve in Figure~\ref{deltaomega} shows $\Delta \Omega$, the
difference between the free-energy density of the mixed phase
(calculated without including the surface and Coulomb energy cost) and
the homogeneous electrically neutral nuclear and CFL phases separated
by a single sharp interface, whichever of the two is lower. The mixed
phase has lower bulk free energy, so $\Delta\Omega$, plotted in
Figure~\ref{deltaomega}, is negative.  The remaining curves in
Figure~\ref{deltaomega} show the sum of the bulk free-energy difference
$\Delta\Omega$ and $(E^S+E^C)$, and the surface and Coulomb energy cost of
the mixed phase calculated using Equation~\ref{sandccost} for droplets, rods
and slabs for three different values of $\sigma_{\rm QCD}$.  Careful
inspection of the figure reveals that for any value of $\sigma_{QCD}$,  
the mixed phase is described as a function of increasing density by a 
progression from drops to rods to slabs of CFL matter within nuclear 
matter to slabs to rods to drops of nuclear matter within CFL matter.  
This is the same progression of geometries seen in
the inner crust of a neutron star \cite{pasta} or in the mixed phase
at a first-order phase transition between nuclear matter and unpaired quark
matter \cite{Heiselberg:1993dx} or hadronic kaon condensate 
\cite{Glendenning:1998zx}.  We have also checked that for
$\sigma_{\rm QCD}=10$ and $40$~MeV/fm$^2$, with the mixed-phase geometry at any $\chi$ taken to be favored, the regions
of both the rarer and more common phases ($r_0$ and its suitably
defined counterpart) are always less than $5-6$ fm. In general,
uniform regions of charge can exist only on length scales small
compared to the Debye screen length. The Debye screening length in the
quark and hadronic phase are typically  $5-10$ fm.  When
the size of the charged regions becomes comparable to the Debye
screening length it becomes important to account for spatial variations
of the charge density. This will influence the surface and Coulomb
energy estimates presented in Equation~\ref{sandccost}. A detailed
discussion of the importance of these finite size effects is presented
in References \cite{Norsen:2000wb} and \cite{Voskresensky:2002hu}. 

For any given $\sigma_{\rm QCD}$, the mixed phase has lower free
energy than homogeneous neutral CFL or nuclear matter whenever one of
the curves in Figure~\ref{deltaomega} for that $\sigma_{\rm QCD}$ is
negative.  We see that much of the mixed phase will survive if
$\sigma_{\rm QCD}\simeq 10~\MeV/ \fm^2$, whereas for $\sigma_{\rm QCD}
\gtrsim 40~\MeV/\fm^2$ the mixed phase is not favored for any $\mu$.
This means that if the QCD-scale surface tension $\sigma_{\rm QCD}
\gtrsim 40~\MeV/\fm^2$, a single sharp interface will be favored. The
interface is characterized by a bipolar charge distribution, resulting
in an intense electric field that ensures that the electric charge
chemical potential is continuous across it (see Ref.~\cite{Alford:2001zr} for details).

\section{CONFRONTING OBSERVABLES}
         \label{sec:section5}

Observations relating to the structure and the evolution of compact stars furnish 
complementary information about the interior state. 
The former provides information about the EoS of dense matter and 
probes the high-energy or short-distance aspects, whereas the latter provides
information about the low-lying excitation spectrum. 
This complementarity potentially allows us to probe the phase structure of matter 
at extreme density, as phase transitions can drastically alter the low-lying 
spectrum without strongly influencing the EoS or vice versa. 
For example, nucleon superfluidity and superconductivity dramatically alter the 
excitation spectrum (exponentially suppressing neutrino emission rates) but 
play no role in the EoS.
In contrast, hyperons strongly soften the EoS but may not affect the linear 
response properties to the same degree.  

\subsection{Bulk Properties and Structure}
            \label{sec:Bulk} 

Mass and radius are certainly the most natural observables that probe the EoS 
of the dense interior. 
Given the relationship between pressure and energy density, the general relativistic 
equation of hydrostatic equilibrium (termed the Tolman-Oppenheimer-Volkoff, TOV, equation) 
uniquely determines the structure of the star, in particular its mass and radius, 
for a given central density \cite{GBOOK}. 
For nonrotating stars, each high-density EoS uniquely specifies a mass-radius curve. 
In the context of neutron star structure, EoSs are characterized as soft or 
stiff relative to each other on the basis of the ratio of pressure to energy density. 
This, however, depends on the density at which the comparison is made. 
The EoS with the larger (smaller) ratio at a specific density is termed stiff (soft).  
An EoS that is soft on average will lead to more compact stars, compared with an EoS that is stiff, 
and reach a smaller maximum mass $M_\mathrm{max}$. 
This is easily understood by noting that the energy density is the source of 
gravity, whereas pressure provides resistance to the gravitational squeeze. 

From model studies it is empirically known that the maximum allowed mass is most 
sensitive to the EoS at the highest (supranuclear) density, whereas the radius is 
sensitive to the EoS in the vicinity of nuclear density \cite{Lattimer:2001}. 
In particular, for stars composed of nucleon matter, the radius appears to be fairly 
sensitive to the density dependence of the nuclear symmetry energy. 
The symmetry energy can potentially be inferred from terrestrial measurements of the 
neutron-skin thickness \cite{Horowitz:2001} and probably from heavy-ion experiments \cite{Li:2006}. 
For a recent review of the role of the nuclear symmetry energy in neutron star structure and 
terrestrial experiment, see Reference~\cite{Steiner:2005}. 
These experiments likely constrain models of the nuclear EoS that 
currently differ significantly at both low and high density. 
For example, the mean-field models \cite{GBOOK} tend to be stiff at low density and soft 
at high density compared with variational calculations such as those obtained by 
Akmal et al. \cite{Akmal:1998cf}. 

Phase transitions to novel phases at supranuclear density typically result in a 
softening of the EoS. 
This arises because these transitions furnish new degrees of freedom that contribute 
more to the energy than to the pressure. 
For example, a $\Lambda$ hyperon that replaces a neutron at the Fermi surface has 
a much lower momentum and a larger rest mass. 
In general, this is true for all hadronic transitions studied to date. 
Clearly the extent of the softening depends on the details of both the nuclear and exotic EoSs.
To successfully employ a mass-radius constraint that infers 
the phase structure of the dense interior, softening should be significant.  
This quantitative expectation is, however, difficult to assert and may not be true in general.
For example, early studies that employed the naive bag model indicated that the quark-hadron 
transition would lead to significant softening, but recent work, 
which takes into account corrections due to quark-quark correlations and superconductivity, 
suggests that the quark EoS might mimic nuclear behavior \cite{mimic}. 

\begin{figure}[t]
\begin{center}
\centerline{\psfig{figure=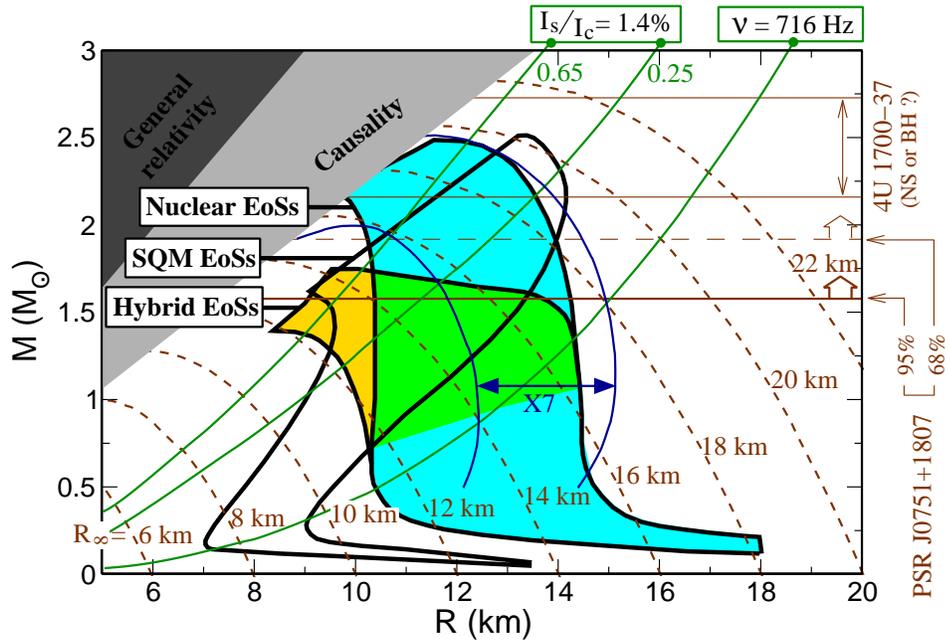,width=13.0cm}}
\end{center}
\caption{Mass-radius diagram: The three generic classes of equations of state
are depicted with observational constraints.}
\label{fig:mr}
\end{figure}

Figure~\ref{fig:mr} depicts the $M-R$ relationship of three generic classes of stars: 
(a) nuclear stars, (b) hybrid stars, and  (c) strange (SQM) stars. 
Here we show bands rather than individual $M-R$ curves for specific models to indicate 
the inherent model dependence of these theoretical predictions.  
As discussed above, the broad range of predicted radii for nucleon EoSs will be narrowed in the
near future owing to neutron-skin thickness and probably also to heavy-ion experiments.
As anticipated in hybrid stars, where there is a phase transition from a nuclear to an exotic 
phase at supranuclear density, softening leads to smaller radii and lower maximum masses. 
In contrast, the most exotic scenario involving absolutely stable SQM overlaps 
the regions accessed by nuclear and hybrid stars for masses in the observed range (1-2 $M_\odot$).
Numerous constraints on the EoS are depicted on the figure.
General relativity requires that the radius be larger than the Schwarzschild radius $R_S$ and 
causality of the EoS requires that the speed of sound $c_s = (dP/d\rho)^{1/2}$ be smaller 
than the speed of light.
This latter constraint is approximately equivalent to $R_\mathrm{max} > 1.5 R_\mathrm{S \; max}$, 
where $R_\mathrm{max}$ and $R_\mathrm{S \; max}$ are the radius and the Schwarzschild radius, 
respectively, of the maximum mass star \cite{LPMY90}.
(Note that the softest nuclear EoSs, resulting from nonrelativistic models, do violate causality
at densities close to the maximum density reached at $M_\mathrm{max}$.)
The other curves in Figure~\ref{fig:mr} refer to various constraints arising from observations:
(a) lower limits on the mass of PSR J0751+1807, at the 68\% and 90\% C.L. \cite{Netal05},
and lower limits on the estimate (68\% C.L.) of the mass of 4U 1700-37 \cite{Clarketal02}, 
which may be a neutron star or a black hole;
(b) the mass-shedding limit on the maximum radius of the fastest known pulsar PSR J1748-2446ad 
\cite{Hessel2006};
(c) 90\% C.L. of the radius at infinity $R_\infty$ of the quiescent LMXB X7 \cite{Rybicki06}
discussed in Section~\ref{sec:qLMXB}; and finally
(d) the minimal radius of glitching pulsars deduced from the constraint 
$I_s/I_c > 1.4\%$, (see Section~\ref{sec:Others}),
assuming the glitch reservoir is the neutron superfluid in the crust.
Converting this moment of inertia constraint into a mass-radius constraint simply requires integrating
the TOV equation from the crust-core boundary to the surface: 
The pressure $P_t$ at this transition point is, however, uncertain, 
with values ranging from $0.25$ up to 0.65 MeV fm$^{-3}$ \cite{LEL99} and the two 
$I_s/I_c$ curve in Figure~\ref{fig:mr} labelled 0.25 and 0.65 
correspond to these two extremes.
For each of the two classes of EoSs, nucleon or hybrid, low $P_t$ EoSs give small radii at low masses,
whereas high $P_t$ give large radii: 
The glitch constraint is hence much more restrictive for EoSs which
are soft at densities around nuclear matter density.

These various constraints shown in Figure~\ref{fig:mr} are already sufficient to draw
some conclusions about the high-density EoS: \\
\begin{enumerate}
\item
The softest range of the hybrid EoSs shown in the figure has a maximum mass above the accuretely
measured mass of the PSR B1913+16 companion, $1.4408 \pm 0.0003 M_\odot$ \cite{WT03}, 
but they are now ruled out at the 90\% C.L. by the new measurement of PSR J0751+1807.
Moreover, all hybrid EoSs are ruled out at the 68\% C.L. by the same measurement.\\
\item
Hybrid EoSs which are soft at low density, i.e., have a low $P_t \sim 0.25$ MeV fm$^{-3}$,
are also strongly disfavored by the glitch $I_s/I_c \simgreater 1.4\%$ constraint. \\
\item
The same EoSs are also in strong disagreement with the $R_\infty$ measurement of X7. 
More reliable measurements of neutron star masses in quiescent LMXBs are needed
to confirm this conclusion and are likely to become available in the near future,
as we discussed in Section~\ref{sec:qLMXB}. \\
\item
The softest nucleon EoSs, which have $P_t \sim 20$--30 MeV fm$^{-3}$, are marginally
consistent with the glitch constraint and need glitching pulsars to have masses
$M \simless 1.4 M_\odot$. \\
\item
Given the broad range of predicted radii for strange stars, none of the present constraints 
can exclude them or make a strong case for their existence.
\end{enumerate}
      
\subsection{Low-Lying Spectrum and Thermal Behavior} 

In contrast to bulk properties, the response properties influence observable aspects such as 
thermal evolution and typically probe energy scales of the order of tens to hundreds of 
keV, and are very sensitive to the low-lying excitation spectrum. 
In some respects, they also tell us about the subtle properties that determine the 
phase structure of dense matter. 

\begin{figure}[t]
\begin{center}
\centerline{\psfig{figure=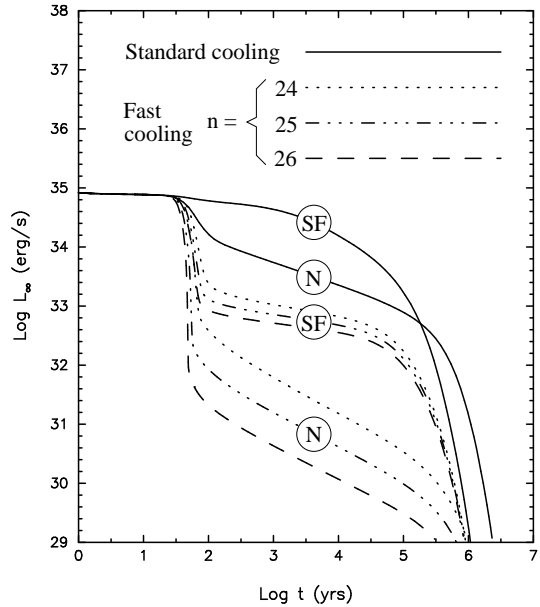,height=8.0cm}}
\end{center}
\caption{Slow, ``standard'', cooling via the modified URCA process
versus fast neutrino cooling from three neutrino emission processes with emissivities
$\epsilon_\nu = 10^n \times T_9^6$ erg cm$^{-3}$ s$^{-1}$.
Trajectories marked as N assume neutrons in the core are normal, whereas
cases marked SF assume neutrons are paired in the $^3P_2$ channel in the whole
core with a $T_c$ of the order of $2$--$3 \times 10^9$ K.
The fast-cooling case with n=24, 25, and 26 corresponds approximately to the
emissivities of a kaon condensate, a pion condensate, and the nucleon direct URCA process,
respectively.
Note that at early times, $t \simless$ 30 yrs, as long as the age of the star is inferior
to the thermal diffusion time scale through the crust, all models result in the same
surface temperature and luminosity.
Neutrino emission from the Cooper pair formation process is not taken into account in these
simple models. 
Figure adapted from Reference \cite{Page1996}.}
\label{fig:Cool-Basic}
\end{figure}

\subsubsection{The Cooling of Isolated Neutron stars} \label{sec:coolingIsolated}

The long-term cooling of neutron stars is the best studied example in this class.
This topic was recently reviewed in detail \cite{Yakovlev:2004,PGW06}
and we present only a short description here.
This cooling is driven by neutrino emission from the core of the star when its temperature is 
above $10^7-10^8$ K or $1-10$ keV. 
The time scale for cooling can be estimated in terms of two key microscopic ingredients, 
namely the neutrino luminosity, $L_\nu$, and the total specific heat, $C_v$,
supplemented by the surface photon luminosity, $L_\gamma$, from a simple energy balance consideration
as follows 
\be
\frac{dE_\mathrm{th}}{dt} = C_v \frac{dT}{dt} = -L_\nu - L_\gamma \;\; ,
\label{eq:cooling}
\ee
where $E_\mathrm{th}$ is the thermal energy content of the star.
For a degenerate Fermi system, the specific heat  $C_v$ is linear in $T$, i.e., 
$C_v \simeq C \cdot T$, where 
the constant $C$ depends on the details of the star's structure.
The neutrino luminosities can be split, in a first approximation, 
into fast or slow processes and can be written as
$L_\nu^\mathrm{fast}  \simeq N^\mathrm{f} \cdot T^6$ or
$L_\nu^\mathrm{slow} \simeq N^\mathrm{s} \cdot T^8$ , respectively.
Finally, $L_\gamma \propto T_e^4$ and because, very roughly,
$T_e \propto T^{1/2}$, one can also write $L_\gamma \simeq S \cdot T^2$.
Owing to their very different $T$ dependences, at early times $L_\nu \gg L_\gamma$, 
whereas when $T$ is sufficiently low the converse is true.
From Equation~\ref{eq:cooling}, the two neutrino cooling time scales for fast and 
slow neutrino emission are
\be
\tau_\nu^\mathrm{fast} = \frac{C}{4N^\mathrm{f} T^4}\approx
\frac{4 \; \mathrm{minutes}}{T_9^4}
\ee
and
\be
\tau_\nu^\mathrm{slow} = \frac{C}{6N^\mathrm{s} T^6} \approx 
\frac{6 \; \mathrm{months}}{T_9^6} \; ,
\ee
respectively.
In obtaining the approximate numerical estimates (4 minutes and 6 months) we have
used typical values of the constants $C$, $N^\mathrm{f}$, and $N^\mathrm{s}$, which are
characteristic of degenerate dense hadronic matter \cite{PGW06}.
For fast cooling, $L_\nu$ dominates $L_\gamma$ until $T \sim 10^6$ K,
whereas for slow cooling this happens when $T \sim 10^8$ K.
Coincidentally, in both cases this turns out to occur at ages of the order of 10$^5$ years. 

In general, exotic phases almost invariably result in fast neutrino emission
(the CFL phase is the only known exception),
and detection of a fast-cooling neutron star would be an argument in 
favor of the presence of one of these phases.
However, because nuclear matter with a high enough proton fraction also results in fast
neutrino emission through the nucleon direct URCA process, such a detection would only be an
indication and not proof of one of these phases.
Moreover, because pairing results in a strong suppression of the neutrino emissivity of all processes
in which the paired component participates, an exotic phase may be present and may not manifest
in the thermal evolution of the star.
We illustrate these considerations in Figure~\ref{fig:Cool-Basic}, in which the time evolution
of the thermal luminosity $L_\infty$ of a model neutron star
for the case of the slow ``standard'' cooling from the modified URCA process is compared with
three different fast neutrino emission processes.
Each scenario is presented in two forms, one in which neutrons are normal
and one in which where they are assumed to be superfluid in the whole core.
The occurrence of fast cooling leads to thermal luminosities, at ages $\sim 10^2$--$10^5$ yrs,
which are about three orders of magnitude lower than in ``standard'' cooling. 
However, when the suppression of the neutrino emissivity by pairing is taken into account,
the differences can be much smaller.
Higher values of $T_c$ can even render fast and slow cooling indistinguishable
\cite{PA92}.
Moreover, once the fast neutrino emission is controlled by pairing gaps, the difference between 
cooling efficiencies of various fast processes becomes much smaller than when they
proceed unhindered, as the three fast-cooling superfluid (SF) curves of 
Figure~\ref{fig:Cool-Basic} show when compared with corresponding curves with normal (N) neutrons.
Considering the plethora of channels for fast neutrino emission when the number of degree
of freedom increases, (a) pion or kaon condensates, (b) hyperons, and (c) deconfined quark matter, 
it appears unlikely that these kinds of studies by themselves will be able to distinguish 
between types of exotica.
Reference \cite{PPLS00} explicitely showed that model stars with only nucleons, 
or nucleons with hyperons,
or nucleons with deconfined quark matter or nucleons with hyperons and deconfined quark matter
can lead to essentially indistinguishable cooling trajectories once uncertainties about the size
of the various pairing gaps are taken into account.
With respect to strange stars, if the SQM surface is covered by a thin envelope of normal
nuclear matter, their thermal evolution would be similar to the one of the more mundane
neutron stars \cite{PGW06,Blaschke:2001} and only in the case of a bare 
quark surface would their thermal evolution be drastically different \cite{PU:2002}.

\begin{figure}[t]
\begin{center}
\centerline{\psfig{figure=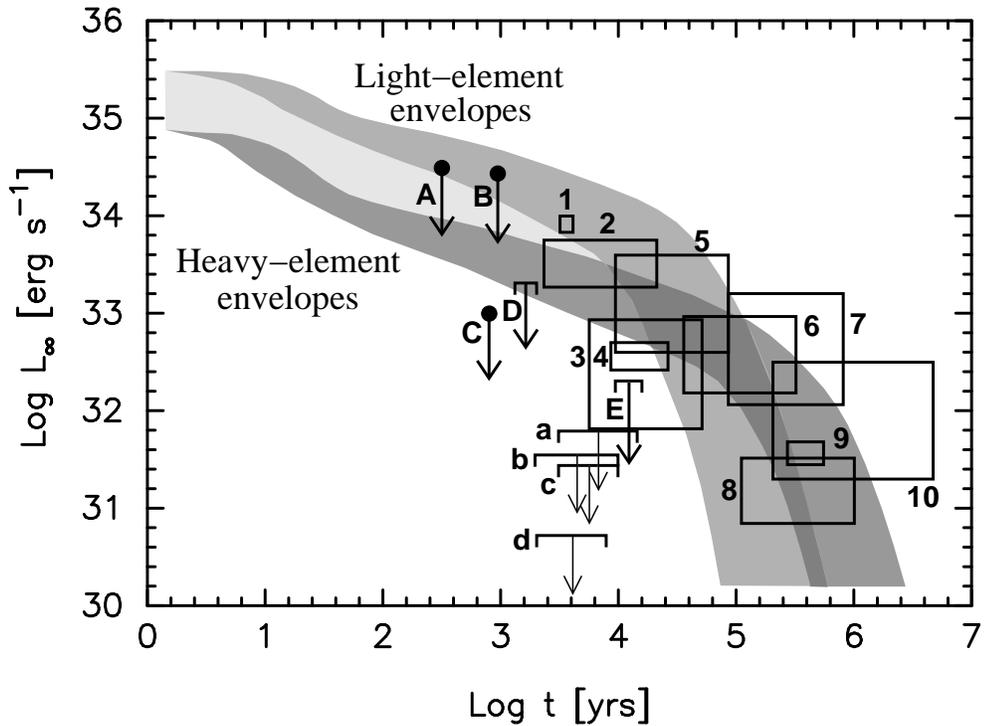,width=13.0cm}}
\end{center}
\caption{Comparison of the prediction of the minimal cooling paradigm
with the observational data presented in Figures~\ref{fig:Cool_Data_Upper}
and \ref{fig:Cool_Data_Good}.
Figure adapted from Reference \cite{PLPS04}.}
\label{fig:minimal}
\end{figure}

Given these intricacies, Page et al. \cite{PLPS04} introduced the minimal cooling scenario as a
paradigm in which all exotic phases are assumed to be absent and the proton fraction is also
assumed to be below the direct URCA threshold.
This paradigm is a benchmark against which data are compared so that observations that cannot
be accommodated within the predictions of this scenario are serious evidence
for the presence of some fast-cooling agent.
Minimal cooling is not naive cooling: 
It considers all others processes that can affect the 
thermal evolution of the star and are based on very standard physics.
Two standard ingredients can influence on the star's cooling significantly.
The first is the occurrence of neutron superfluidity and/or proton superconductivity,
which suppresses the specific heat and the neutrino emission by the modified URCA and 
bremsstrahlung processes but also opens the new neutrino emission channels owing to the constant
formation and breaking of Cooper pairs.
The net effect of pairing can hence  be either a reduction or an enhancement of $L_\nu^\mathrm{slow}$.
The second important ingredient is the chemical composition of the uppermost layers of the
star, the envelope, which controls the photon luminosity $L_\gamma$: 
A light-element envelope, H, He, C, or O,
is less insulating than a heavy-element, iron-like envelope and results in a larger
$L_\gamma$ and $T_e$ for a given internal temperature.
A light-element envelope has the effect of making the star brighter during the neutrino cooling
era because at that time $L_\gamma$ is negligible compared to $L_\nu$, and the surface temperature 
$T_e$ just follows the evolution of the internal temperature while accelerating the cooling
owing to its higher $L_\gamma$ during the photon cooling era.
In Figure~\ref{fig:minimal}, the predictions of the minimal cooling paradigm are displayed 
and compared with the observational data presented in Section~\ref{sec:cooldata}.
Models with a light-element and heavy-element envelope are separated, and for each class the
spread in prediction is owing to the uncertainty in the size of the neutron and proton pairing gaps.
Note that the luminosities at ages 10$^2$--10$^5$ yrs can be quite lower than those of the
``standard'' cooling of Figure~\ref{fig:Cool-Basic} because of the enhanced neutrino emission
from the Cooper pair formation process.
Overall the agreement between the theory and the observational data is quite good.
Two objects, PSR J0205+6449 (in SNR 3C58) and RX J0007.0+7302, nevertheless
have upper limits below these predictions and are good candidates for the occurrence of
fast neutrino emission, controlled by pairing, but they may also be interpreted
as evidence for medium-enhanced modified URCA process \cite{Blaschke:2004}.
The four upper limits, marked a, b, c, and d in Figure~\ref{fig:minimal},
from Kaplan et al.'s \cite{Kaplan2004} search, would, however, be definitive evidence for
fast cooling if they can be proven to refer to neutron stars and not black holes.
Similar conclusions have been reached by the St. Petersburg group \cite{Gusakov},
which has developed a complementary version of minimal cooling.

\subsubsection{The Thermal Behavior of Neutron Stars in Low-Mass X-Ray Binaries} 
\label{sec:coolingSXRT}

We presented in Section~\ref{sec:SXRT} observations of neutron stars in LMXBs
that are undergoing transient phases of accretion 
separated by periods of quiescence.
Brown et al.'s \cite{BBR98} mechanism of heating of the star
provides a simple explanation for the hot thermal spectrum observed in quiescence
in the cases of Aql X-1, 4U 1608-522, and CX 1, but the results of Table~\ref{tab:SXRT} clearly
show that many systems have quiescent luminosities much below the prediction of this model.
A natural explanation for this discrepancy is the presence of fast neutrino cooling
in the core of these too-cold neutron stars \cite{CGPP01}.
This result condradicts sharply with the fact that most isolated cooling
neutron stars have thermal luminosities in agreement with the results of the minimal
cooling scenario, i.e., they show no conclusive evidence for the occurrence of fast neutrino emission. 
However, remembering that LMXBs are very long-lived systems in which the neutron star
may accrete a significant part of a solar mass, 
these cold neutron stars may simply be massive enough to be above the threshold for 
fast neutrino emission, whereas most of the isolated cooling neutron stars are below this threshold.
These results would be in good agreement with the expected mass range of neutron stars
emerging from core collapse supernovae as illustrated in Figure~\ref{fig:IMF} and are corroborated
by the measured masses of PSRs in PSR+NS systems exhibited in Figure~\ref{fig:Masses},
whereas the measured masses of PSR+WD systems, offsprings of LMXBs, confirm the existence
of more massive neutron stars.

\subsection{Emerging Trends and Future Prospects}

\subsubsection{Limiting Spin Frequency}

The first, absolute, limiting frequency for a fast-spinning pulsar is the mass-shedding limit
$\Omega_\mathrm{MS}$, which we used as a constraint in Figure~\ref{fig:mr}, from the
fastest known pulsar PSR J1748-2446ad.
However, gravitational radiation reaction-driven instabilities such as the r-mode 
instability can result 
in a more stringent limit on the spin frequency. These modes generate gravity waves (which dissipate 
angular momentum) that, instead of damping, amplify mode amplitude \cite{Rmode}. 
Viscosity is then the main source of damping and, depending on the bulk and shear viscosity of 
dense matter, different limiting spin frequencies are possible. 
At very low temperature, damping is due to the shear viscosity, whereas at very high temperature,
it is due to the bulk viscosity.
For these extreme temperatures, r-modes are effectively damped.
At intermediate values of temperature, 
depending on the bulk viscosity and its temperature dependence, a limiting frequency as low as 
 50\% of $\Omega_\mathrm{MS}$ is possible. In LMXBs where the neutron stars are spun up to 
millisecond periods, a population study shows a sharp drop in the pulsars with spin frequency 
$\Omega \simgreater 730$ Hz \cite{Deepto}. 
If this limiting behavior is indeed due to gravity waves from the r-mode instability, 
it provides useful information about the viscous damping rate in the star, which can 
be calculated for different EoSs for the neutron star interior. 
For example, a strange quark star made entirely of CFL quark matter would be incompatible 
with these observations because all low-energy weak interaction processes that can contribute 
to the bulk viscosity are exponentially suppressed by $\exp(-\Delta/k_B T)$, where $\Delta$ 
is the pairing gap and $T$ is the temperature \cite{Madsen}. 

\subsubsection{Galactic Supernova Neutrinos}

The enormous gravitational binding energy, $E_\mathrm{G}\simeq GM^2/R\simeq 3\times10^{53}$ ergs,  
gained during a core collapse supernova is stored inside the newly born proto-neutron star as 
thermal energy of the matter components and thermal and degeneracy energy of neutrinos. 
This energy then leaks out on a time scale determined by the rate of diffusion of neutrinos in 
the dense core \cite{burr:86}. 
The detection of  $\sim 20$ neutrinos from SN1987A is testimony to this theoretical expectation. 
Current estimates indicate that we should see $\sim10,000$ events in SuperKamiokande and 
$\sim1000$ events  in the Sudbury Neutrino Observatory (SNO) from a supernova at the center of our 
galaxy (distance=$8.5$ kpc) \cite{Burrows:1992cy}.
Detection would provide detailed information about the temporal structure of the 
neutrino signal and thereby probe the properties of the dense inner core. 
Preliminary studies of the role of phase transitions in the evolution of proto-neutron stars 
and the supernova neutrino signal exist and show that the late time signal is 
sensitive to the high-density physics. 
In particular, a dramatic consequence of the  transition to a very soft high-density EoS 
would be a delayed collapse to a black hole on the neutrino diffusion time scale \cite{SNnus}. 
Less dramatic changes to the neutrino luminosity can be detected for a galactic supernova and 
could potentially probe significant changes to the neutrino mean free path that occur as a result 
of a phase transition. Like neutrino emissivities, the neutrino opacity of dense matter is 
sensitive to the phase structure and low-lying excitation spectrum of dense matter \cite{NuOpac}. 

\subsubsection{Gravity Waves}
           
The detection of gravity waves \cite{GW} from a binary system in 
Laser Interferometer Gravitational-Wave Observatory-like detectors, which can measure 
frequencies up to 1 kHz, can measure the chirp mass $M_{ch}=\mu^{3/5} ~M_T^{2/5}$, where $\mu$ 
is the reduced mass and $M_T$ is the total mass of the binary \cite{Chirp}. 
Perhaps more interestingly, during the inspiral the compactness $M/R$ 
is also accessible and will thereby provide an estimate of the star radius \cite{Faber:2002zn}.  
In addition, a first-order phase transition occurring in the inner core of a neutron star could 
lead to a density discontinuity if the surface tension is large. 
In Reference \cite{GWave}, the authors show that this would affect the frequency spectrum of the 
non-radial oscillation modes in two ways. 
First, it would produce a softening of the EoS, leading to more compact 
equilibrium configurations and changing the frequency of the fundamental and pressure 
modes of the neutron star. 
Second, a new nonzero-frequency g-mode would appear in association with each discontinuity. 
If these modes are excited in either a phase transition-induced mini-collapse or a 
glitch (binary neutron star mergers are unlikely to excite these high-frequency modes), 
they may be observable in the next-generation gravity wave detectors.  
The role of the EoS in binary neutron star mergers is also potentially interesting 
and largely unexplored. 
Reference \cite{SQMGwave} finds that if there is time for stable mass transfer to
occur during the merger, it could provide information about the mass-radius 
diagram and potentially distinguish between normal stars and strange quark stars.

 \section{CONCLUSIONS}
          \label{sec:section6}

To infer the phase structure and properties of matter at extreme density through observations of compact stars is a compelling but difficult task. Despite the rather rudimentary state of quantitative theoretical calculations in describing dense matter, the observational data are already providing guidance to and hinting at possible tensions between different models. In particular the prospect for a firm mass measurement of a heavy neutron star (or a low-mass black hole) and accurate radius measurements are excellent. Such measurements will directly confirm or rule out the possibility for a strong phase transition to a soft EoS in the vicinity of nuclear density, and presently available values already seem to disfavor a soft EoS. 
Observations of cooling neutron stars and the realization that superfluidity and superconductivity play a dominant role in neutrino cooling of the core has sparked renewed interest in theoretical studies of pairing correlations in dense matter. Furthermore, the connection between seemingly different observables such as long-term neutron star cooling, quiescent luminosity observed in LMXBs, and superbursts are indicating new trends that point to a diversity in the thermal evolution of neutron stars, with growing evidence in favor of the occurence of some fast neutrino emitting process(es), at least in heavy, $>1.4 M_\odot$, neutron stars.
These developments call for a more global approach in the theory and simulations to confront different observables within a unified framework so that the sensitivity to the underlying dense matter physics can be properly assessed. 

Theoretical progress in describing dense nuclear matter and new insights into using terrestrial experiments on nuclei to better constrain 
EoS models provide compelling reasons to expect a quantitative description of neutron-rich matter for the densities of interest. Throughout this review we highlighted important qualitative differences and possible similarities between nucleon matter and other exotic phases. Minimizing the theoretical uncertainty for the predictions of the nucleon matter is crucial to confront observables. For example, a precise determination of the density dependence of the nuclear symmetry energy can reduce the uncertainty simultaneously in radii and cooling rates predicted for the different nuclear EoSs.   The situation with exotic states of matter is less clear. Although softening and enhanced cooling appear to be common trends, there are important exceptions and quantitative estimates are often very model dependent. Nonetheless, as mentioned above, radius measurement smaller than 10 km or the discovery of low-mass black hole candidate with  mass $ \simless 2 M_\odot$  would be very suggestive of a phase transition. Better experimental or theoretical constraints on the hyperon-nucleon and kaon-nucleon interactions are necessary to ascertain if novel hadronic phases are relevant.  Quark matter remains an especially interesting possibility. Color superconductivity is relevant, especially for the calculation of neutrino processes and cooling. However, an understanding of how a finite strange quark mass affects the phase structure is needed because this can result in vastly different low energy properties.

\section*{Acknowledgments}
D.P. is partially supported by Universidad Nacional Aut\'onoma de M\'exico
Direcci\'on General de Asuntos del Personal Acad\'emico grant number IN119306-3.
S.R. is supported by the U.S. Department of Energy contract number W-7405-ENG-36.
The authors thank C. Fryer, B. Link, M. van Kerkwijk, G. G. Pavlov, and V. E. Zavlin,
for discussions about this work and/or permission to reprint figures from 
their published works.
We would also like to thank M. Alford for his
comments and suggestions.




\begin{thebibliography}{99}

\bibitem{TC99}
Thorsett SE, Chakrabarty D.
{\em Astrophys. J.} 512:288 (1999)

\bibitem{S04}
Stairs IH.
{\em Science} 304:547 (2004)

\bibitem{Netal05}
Nice DJ, et al. 
{\em Astrophys. J.} 634:1242 (2005)

\bibitem{Setal05}
Splaver EM, et al.
{\em Astrophys. J.} 620:405 (2005)

\bibitem{Retal05}
Ransom SM, et al.
{\em Science} 307:892 (2005)
 
\bibitem{Jetal05}
Jacoby BA, et al.
{\em Astrophys. J.} 629:L113 (2005)

\bibitem{vdMKvKvdH05}
van der Meer A, et al.
In {\em Interacting Binaries: Accretion, Evolution, and Outcomes},
AIP Conf. Proc. 797:623 (2005)

\bibitem{vKvPZ95}
van Kerkwijk MH, van Paradijs J, Zuiderwijk EJ.
{\em Astron. Astrophys.} 303:497 (1995)

\bibitem{Betal01}
Barziv O, et al.
{\em Astron. Astrophys.} 377:925 (2001)

\bibitem{OK99}
Orosz JA, Kuulkers E.
{\em Mon. Not. R. Astron. Soc.} 305:132 (1999)

\bibitem{DAetal05}
D'Avanzo P, et al.
{\em Astron. Astrophys.} 444:905 (2005)

\bibitem{MDCMP05}
Mu\~{n}oz-Darias T, Casares J, Martinez-Pais IG.
{\em Astrophys. J.} 635:502 (2005)

\bibitem{Setal03}
Shahbaz T, et al. 
{\em Astrophys. J.} 585:433 (2003)

\bibitem{Clarketal02}
Clark JS, et al.
{\em Astron. Astrophys.} 392:909 (2002)

\bibitem{JSNvdK05}
Jonker PG, Steeghs D, Nelemans G, van der Klis M.
{\em Mon. Not. R. Astron. Soc.} 356:621 (2005)

\bibitem{ST84}
Shapiro S. L., Teukolsky S. A. 1984.
{\em Black Holes, White Dwarfs, and Neutron Stars: the Physics of Compact Objects},
New York: John Wiley

\bibitem{TWW96}
Timmes FX, Woosley SE, Weaver TA.
{\it Astrophys. J} 457:834 (1996);
Woosley SE, Heger A, Weaver TA. {\em Rev. Mod. Phys.} 74: 1015 (2002)

\bibitem{FK01}
Fryer CL, Kalogera V.
{\em Astrophys. J.} 554: 548 (2001)




\bibitem{MMcM83}
Morrison R, McCammon D.
{\em Astrophys. J.} 270:119 (1983)

\bibitem{R87}
Romani RW.
{\em Astrophys. J.} 313:718 (1987)

\bibitem{ZP02}
Zavlin VE, Pavlov GG.
In {\em Neutron Stars, Pulsars, and Supernova Remnants},
ed. W Becker, H Lesch,  J Tr\"umper, p. 263.
Garching, M\"unchen: Max-Plank-Inst.,
MPE Report 278 (2002) astro-ph/0206025

\bibitem{ZPS96}
Zavlin VE, Pavlov GG, Shibanov Yu A.
{\em Astron. Astrophys.} 315:141 (1996);

\bibitem{NoBSpectra}
G\"{a}nsicke BT, Braje TM, Romani RW.
{\em Astron. Astrophys.} 386:1001 (2002);
Werner K, Deetjen J.
In {\em Pulsar Astronomy - 2000 and Beyond}, 
ed. M Kramer, N Wex, N Wielebinski, p. 623.
San Francisco: Astron. Soc. Pac., IAU Symp. 202 (2000)
astro-ph/0007094

\bibitem{PSVZ94}
Pavlov GG, Shibanov YuA, Ventura J, Zavlin VE.
{\em Astron. Astrophys.} 289:837 (1994)

\bibitem{RRC97-M92}
Miller MC.
{\em Mon. Not. R. Astron. Soc.} 255:129 (1992)
Rajagopal M, Romani RW, Miller MC.
{\em Astrophys. J.} 479:347 (1997)

\bibitem{MagHHE}
Shibanov IuA., Zavlin VE, Pavlov GG, Ventura J.
{\em Astron. Astrophys.} 266:313 (1992);
Ho WCG, Lai D.
{\em Mon. Not. R. Astron. Soc.} 327:1081 (2001);
Lloyd DA. astro-ph/0303561

\bibitem{Lai01}
Lai D.
{\em Rev. Mod. Phys.} 73:629 (2001)

\bibitem{MagSurf}
Turolla R, Zane S, Drake JR.
{\em Astrophys. J.} 603:265 (2004);
Perez-Azorin JF, Miralles JA, Pons JA.
{\em Astron. Astrophys.} 433:275 (2005);
van Adelsberg M, et al.
{\em Astrophys. J.} 628:902 (2005)

\bibitem{GH83}
Greenstein G, Hartke GJ. 
{\em Astrophys. J.} 271:283 (1983)

\bibitem{P95}
Page D. 
{\em Astrophys. J.} 442:273 (1995)

\bibitem{PGW06}
Page D., Geppert U., Weber F.
{\em Nucl. Phys. A.} in press (2006) astro-ph/0508056

\bibitem{GKP06}
Geppert U, Kuecker M, Page D.
{\em Astron. Astrophys.} in press (2006) astro-ph/0512530

\bibitem{PLPS04}
Page D, Lattimer JM, Prakash M, Steiner AW.
{\em Astrophys. J. Suppl.} 155:623 (2004)


\bibitem{Weisskopf2004}
Weisskopf MC, et al.
{\em Astrophys. J.} 601:1050 (2004)

\bibitem{Kaplan2004}
Kaplan DL, et al.
{\em Astrophys. J. Suppl.} 153:269 (2004)

\bibitem{HaberlReview}
Haberl F.
{\em Adv. Space Res.} 33:638 (2002) astro-ph/0401075


\bibitem{Pons:2002}
Pons JA, et al.
{\em Astrophys. J.} 564:981 (2002);
Truemper JE, Burwitz V, Haberl F, Zavlin VE.
{\em Nucl. Phys. Suppl.} 132:560 (2004)

\bibitem{QuarkStar}
Drake JJ, et al.
{\em Astrophys. J.} 572:996 (2002)

\bibitem{NoQuarkStar}
Braje TM, Romani RW.
{\em Astrophys. J.} 580:1043 (2002);
Prakash M, Lattimer JM, Steiner AW, Page D
{\em Nucl. Phys. A} 715:835 (2003)


\bibitem{RBBPZ2002}
Rutledge RE, et al. 
{\em Astrophys. J.} 577:346 (2002)

\bibitem{Heinke2003}
Heinke CO, et al. 
{\em Astrophys. J.} 598:501 (2003)

\bibitem{Gendre2003a}
Gendre B, Barret D, Webb NA.
{\em Astron. Astrophys.} 400:521 (2003)

\bibitem{Gendre2003b}
Gendre B, Barret D, Webb N.
{\em Astron. Astrophys.} 403:L11 (2003)

\bibitem{Rybicki06}
Heinke CO, Rybicki GB, Narayan R, Grindlay JE.
{\em Astrophys. J.} 644:1103 (2006)


\bibitem{XrayNovae}
Chen W, Shrader R, Livio M.
{\em Astrophys. J.} 491:312 (1997);
Tanaka Y, Shibazaki N
{\em Annu. Rev. Astron. Astrophys.} 34:607 (1996)

\bibitem{Osaki1996}
Osaki Y.
{\em Publ. Astron. Soc. Pac.} 108:39 (1996)

\bibitem{Lasota2001}
Lasota J-P.
{\em New. Astron. Rev.} 45:449 (2001)

\bibitem{HZ}
Haensel P, Zdunik JL.
{\em Astron. Astrophys.} 227:431 (1990);
{\em Astron. Astrophys.} 404:L33 (2003)

\bibitem{BBR98}
Brown EF, Bildsten L, Rutledge RE.
{\em Astrophys. J.} 504:L95 (1998)

\bibitem{CSL97}
Chen W, Shrader CR, Livio M.
{\em Astrophys. J.} 491:312 (1997)

\bibitem{Cackett-etal05}
Cackett EM, et al.
{\em Astrophys. J.} 620:922 (2005)

\bibitem{Wetal03}
Wijnand R, et al. 
{\em Astrophys. J.} 594:952 (2003)

\bibitem{Wetal05b}
Wijnands R, et al.
{\em Astrophys. J.} 618:883 (2005)

\bibitem{Cetal02}
Campana S, et al.
{\em Astrophys. J.} 575:L15 (2002)

\bibitem{Jetal04}
Jonker PG, et al.
{\em Mon. Not. R. Astron. Soc.} 354:666 (2004)

\bibitem{Wetal05}
Wijnands R, et al.
{\em Astrophys. J.} 619:492 (2005)

\bibitem{Campana-etal05}
Campana S, Ferrari N, Stella L, Israel GL.
{\em Astron. Astrophys.} 434:L9 (2005)

\bibitem{Wetal01}
Wijnands R, et al.
{\em Astrophys. J.} 560:L159 (2001)

\bibitem{Cackett-etal06}
Cackett EM, et al.
{\em Mon. Not. R. Astron. Soc.} 369:407 (2006)

\bibitem{Tomsick-etal05}
Tomsick JA, Gelino DM, Kaaret P.
{\em Astrophys. J.} 635:1233 (2005)

\bibitem{Tomsick-etal04}
Tomsick JA, et al.
{\em Astrophys. J.} 610:933 (2004)

\bibitem{CGPP01}
Colpi M, Geppert U, Page D, Possenti A.
{\em Astrophys. J.} 548:L175 (2001)



\bibitem{vanderKlis:2000}
van der Klis M.
{\em Annu. Rev. Astron. Astrophys.} 38:717 (2000)

\bibitem{Cottam:2002}
Cottam J, Paerels F, Mendez M.
{\em Nature} 420:51 (2002)

\bibitem{BurstsReviews}
Lewin WHG, van Paradijs J, Taam RE.
{\em Space Sci. Rev.} 62: 223 (1993);
Lewin WHG.
In {\em X-ray Binaries}, ed. WHG Lewin, J van Paradijs, EPJ van den Heuvel, p. 175.
Cambridge: Cambridge Univ. Press (1995)

\bibitem{Superbursts}
Kuulkers E.
{\em Nucl. Phys. B Proc. Supl.} 132:466 (2004);
Strohmayer T, Bildsten L. 
In {\em Compact Stellar X-Ray Sources}, 
ed. WHG Lewin, M van der Klis.
Cambridge: Cambridge University Press (2006) astro-ph/0301544

\bibitem{Brown:2004}
Brown EF.
{\em Astrophys. J.} 614: L57 (2004);
Cumming A, Macbeth J, in't Zand JMM, Page D.
{\em Astrophys. J.} 646:429 (2006)


\bibitem{LEL99}
Link B, Epstein RI, Lattimer JM.
{\em Phys. Rev. Lett.} 83:3362 (1999)

\bibitem{Hessel2006}
Hessel JWT, et al.
{\em Science} 311:1901 (2006) 

\bibitem{LP:2004}
Lattimer JM, Prakash M.
{\em Science} 304:536 (2004)

%
%

\bibitem{Heiselberg:2000dn}
Heiselberg H, Pandharipande VR.
{\em Annu. Rev. Nucl. Part. Sci.} 50:481 (2000)

\bibitem{Akmal:1998cf}
Akmal A, Pandharipande VR, Ravenhall DG.
{\em Phys. Rev. C} 58:1804 (1998)

\bibitem{CarlsonGFMC}
Carlson J, ~Morales JJ, Pandharipande VR, Ravenhall DG,
{\it Phys. Rev. C} 68: 025802 (2003)

\bibitem{Bethe:1971}
Bethe HA,
{\it Annu. Rev. Nucl. Part. Sci.} 21:93 (1971).

\bibitem{BHF}
Baldo M, ed.
{\it International Review of Nuclear Physics, Vol. 8.
Nuclear Methods and the Nuclear Equation of State} 
Singapore: World Scientific (1999);
Baldo M, Bombaci I, Burgio GF.
{\it Astron. Astrophys.} 328:274 (1997)

\bibitem{Nijmegen:1994} Stoks VGJ, Klomp RAM, Terheggen CPF, de Swart JJ, {\it Phys. Rev. C.} 49:2950 (1994)  

\bibitem{3Body} 
Pudliner BS, et al. 
{\it Phys. Rev. C.} 56:1720 (1997)  

\bibitem{EFT}
 Beane SR, et al. 
arXiv:nucl-th/0008064 (2000)

\bibitem{Vlowk}
  Bogner SK, Kuo TTS, Schwenk A.
 {\it  Phys. Rept.} 386:1 (2003)

\bibitem{Bogner:2005sn}
  Bogner SK, Schwenk A, Furnstahl RJ, Nogga A.
{\it Nucl.\ Phys.\ A} 763:59 (2005)


\bibitem{Furnstahl:2003cd}
Furnstahl RJ,
{\em Lect. Notes Phys.} 641:1 (2004)

\bibitem{Walecka:qa}
Walecka JD.
{\em Annals Phys.} 83:491 (1974)

\bibitem{Serot:1984ey}
Serot BD, Walecka JD.
{\em Adv. Nucl. Phys.} 16:1 (1986)

\bibitem{Boguta:xi}
Boguta J, Bodmer AR.
{\em Nucl. Phys. A} 292:413 (1977)

\bibitem{GBOOK}
Glendenning N. K., {\it Compact Stars, Nuclear Physics, Particle Physics,
and General Relativity }, (Springer-Verlag, New York, 1997)

\bibitem{PREX}
Horowitz CJ, Pollock SJ, Souder PA, Michaels R,
 {\it Phys. Rev. C} 63:025501 (2001);
 Horowitz CJ, Piekarewicz J,
 {\it Phys. Rev. Lett.} 86:5647 (2001)
 
\bibitem{HI}
 Danielewicz P, Lacey R, Lynch WG,
  {\it Science} 298:1592 (2002);
  B.~A.~Li and L.~W.~Chen,
  {\it Phys. Rev. C} 72:064611 (2005);
 Steiner AW, Li BA,
  {\it Phys. Rev. C} 72:041601 (2005)
  
  
\bibitem{Lattimer:1991}
Lattimer JM, Pethick CJ, Prakash M,  Haensel P, 
{\it Phys. Rev. Lett} 66:2701 (1991)

\bibitem{Bardeen:1957mv}
Bardeen J, Cooper LN, Schrieffer JR.
{\em Phys. Rev.} 108:1175 (1957)

\bibitem{SFReview}
Lombardo U, Schulze H-J.
{\em Lect.Notes Phys.} 578:30 (2001)
Dean DJ, Hjorth-Jensen E.
{\em Rev. Mod. Phys.} 75:607 (2003);

\bibitem{large3p2} 
Jackson AD, Krotscheck E, Meltzer DE, Smith RA.
{\it Nucl. Phys. A} 386:125 (1982); 
Pethick CJ, Ravenhall DG 
{\it Ann. N. Y. Acad. Sci.} 647:503 (1991)

\bibitem{small3p2} 
Schwenk A, Friman B.
{\it Phys. Rev. Lett.} 92:082501 (2005) 

\bibitem{Flowers:1976} 
Flowers E, Ruderman M, Sutherland P. 
{\it Astrophys. J} 205:541 (1976)
 
\bibitem{Voskresensky:1986} 
Voskresensky DN, Senatorov AV,.
{\it Sov. Phys. JETP Lett.} 63:885 (1986) ; 
{\it Sov. J. Nucl. Phys. A} 45:411 (1987) 

\bibitem{frim:79}
Friman B, Maxwell OV.
{\em Astrophys. J.} 232:541 (1979)

\bibitem{hanh:00}
Hanhart C, Phillips DR, Reddy S.
{\em Phys. Lett. B} 499:9 (2001)

\bibitem{Voskresensky:2001}
  Voskresensky DN ,
  {\it Lect. Notes Phys.} 578: 467 (2001)

\bibitem{Migdal:1990vm}
Migdal AB, Saperstein EE, Troitsky MA, Voskresensky DN,
{\it Phys. Rept.} 92:179 (1990)

  \bibitem{Schwenk:2003pj}
  Schwenk A, Jaikumar P, Gale C,
{\it Phys. Lett. B} 584: 241 (2004)



%
%


  
\bibitem{Schaffner-Bielich:2000yj}
 Schaffner-Bielich J, Hanauske M, Stoecker H, Greiner W. {\it Phys. Rev. Lett} 89:171101 (2002)  

\bibitem{hypQCD}
  Beane SR, Bedaque PF,  Parreno A, Savage MJ.
  {\it Nucl. Phys. A} 747:55 (2005)

\bibitem{Rijken:1998yy}
Rijken TA, Stoks VGJ, Yamamoto Y.
{\it Phys. Rev. C} 59:21 (1999)


\bibitem{Millener:hp}
Millener DJ, Dover CB,  Gal A.
{\em Phys. Rev. C} 38:2700 (1988)
  
\bibitem{Glendenning:1991es}
Glendenning NK, Moszkowski SA.
{\em Phys. Rev. Lett.} 67:2414 (1991)

\bibitem{Baldo:1998hd}
Baldo M, Burgio GF, Schulze HJ.
{\it Phys.\ Rev.\ C} 58:3688 (1998).


\bibitem{Prakash:1992}
Prakash M, Prakash M, Lattimer JM, Pethick CJ.
 {\it Astrophys. J.} 390:L77 (1992) 

\bibitem{Friedman:hx}
Friedman E, Gal A, Batty CJ.
{\em Nucl. Phys. A} 579:518 (1994)

\bibitem{pionreview}
Migdal AB.
{\em Rev. Mod. Phys.} 50:107 (1978)

\bibitem{Kaplan:yq}
Kaplan DB, Nelson AE.
{\em Phys. Lett. B} 175:57 (1986)

\bibitem{Ramos:2000dq}
Ramos A, Schaffner-Bielich J, Wambach J.
{\em Lect. Notes Phys.} 578:175 (2001) arXiv:nucl-th/0011003

\bibitem{Carlson:1999rr}
Carlson J, Heiselberg H. Pandharipande VR.
{\em Phys. Rev. C} 63:017603 (2001)

\bibitem{MesonCool}
Campbell DK, Dashen RF, Manassah JT. 
{\it Phys. Rev. D} 12:979 (1975); 
Brown GE, Kubodera K, Page D, Pizzochero P. 
{\it Phys. Rev. D} 37:2042 (1988);
Tatsumi T. {\rm Prog. Theor. Phys.} 80:22 (1988)

\bibitem{Thorsson:1995}
Thorsson V, Prakash M, Tatsumi T, Pethick CJ.
{\it Phys. Rev. D} 52:3739 (1995) 

\bibitem{Bag} 
Chodos A, et al.
{\it Phys. Rev. D} 9:3471 (1974)

\bibitem{Freedman:1977gz}
Freedman B, McLerran LD.
{\em Phys. Rev. D} 17:1109 (1978)

\bibitem{Fraga:2001id}
Fraga ES, Pisarski RD, Schaffner-Bielich J.
{\em Phys. Rev. D} 63:121702 (2001)

\bibitem{Barrois:1977xd}
Barrois BC.
{\em Nucl. Phys. B} 129:390 (1977);
Frautschi SC. 
In {\it Workshop on Hadronic Matter at Extreme Energy Density} 
(Erice, Italy 1978)

\bibitem{BigGap}
Alford MG, Rajagopal K, Wilczek F.
{\em Phys. Lett. B} 422:247 (1998);
Rapp R, Schafer T, Shuryak EV, Velkovsky M.,
{\em Phys. Rev. Lett.} 81:53 (1998)

\bibitem{CFLReview}
Rajagopal K, Wilczek F,
{\em The condensed matter physics of QCD}
hep-ph/0011333;
Alford M G.
{\em Annu. Rev. Nucl. Part. Sci.} 51:131 (2001)

\bibitem{Alford:1998mk}
Alford MG, Rajagopal K, Wilczek F.
{\em Nucl. Phys. B} 537:443 (1999)

\bibitem{CFLmesons}
Casalbuoni R, Gatto R.
{\em Phys. Lett. B} 464:111 (1999);
Son DT, Stephanov MA.
{\em Phys. Rev. D} 61:074012 (2000)
erratum, {\it ibid.} 62:059902 (2000);

\bibitem{Alford:2001zr}
Alford MG, Rajagopal K, Reddy S, Wilczek F.
{\em Phys. Rev. D} 64:074017 (2001)

\bibitem{AlfordReddy:2003}
Alford M, Reddy S.
{\it Phys. Rev. D} 67:074024 (2003)

\bibitem{Lugones:2002}
Lugones G, Horvath JE.
{\it Phys. Rev. D} 66:074017 (2002)

\bibitem{NJL} 
Buballa M.
 {\it Phys.\ Rep.} 407:205 (2005)

 
\bibitem{mimic}
 Alford MG, Braby M, Paris MW, Reddy S.
  {\it Astrophys.\ J.}  629:969 (2005)
 
\bibitem{Alford:1999pa}
Alford MG, Berges J, Rajagopal K.,
{\em Nucl. Phys. B} 558:219 (1999)

\bibitem{Buballa:2001gj}
Buballa M, Oertel M.
{\em Nucl. Phys. A} 703:770 (2002)

\bibitem{Bedaque:2001je}
Bedaque PF, Schafer T.
{\em Nucl. Phys. A} 697:802 (2002)


\bibitem{Kaplan:2001qk}
Kaplan DB, Reddy S.
{\em Phys. Rev. D} 65:054042 (2002)

\bibitem{Alford:2002kj}
Alford M, Rajagopal K. 
{\it J. High Ener. Phys.} 6:031 (2002) 

\bibitem{NJLStrong}
Buballa M, et al.
{\it Phys. Lett. B} 595:36 (2004);
Blaschke D, et al. 
{\it Phys. Rev. D} 72:065020 (2005);
Ruster SB, et al.
{\it Phys. Rev. D} 72:034004 (2005)


\bibitem{CSL}
Schafer T.
  {\it Phys. Rev. D} 62:094007 (2000);
  Schmitt A,
  {\it Phys. Rev. D} 71:054016 (2005);
  Alford MG, Cowan GA.
   (2005) hep-ph/0512104

\bibitem{Rajagopal:2000ff}
Rajagopal K, Wilczek F.
{\it Phys. Rev. Lett.}  86:3492 (2001)

\bibitem{Steiner:2002gx}
Steiner AW, Reddy S, Prakash M.
{\it Phys.\ Rev.\ D}  66:094007 (2002)

  \bibitem{Bedaque}
  Bedaque PF.
  {\it Nucl. Phys. A} 697:569 (2002); 
    Bedaque PF, Caldas H, Rupak G.
  {\it Phys. Rev. Lett.} 91:247002 (2003)

  \bibitem{LOFF} 
  Larkin AI, Ovchinnikov YuN. 
 {\it Sov. Phys. JETP} 20:762 (1965); 
  Fulde P, Ferrell RA. 
  {\it Phys. Rev.} 135:A550 (1964) 
   
  \bibitem{Liu:2002}
  Liu WV, Wilczek F.
  {\it Phys. Rev. Lett.} 90:047002 (2003)

   \bibitem{gapless2SC}
  Shovkovy I, Huang M.
  {\it Phys. Lett. B} 564:205 (2003)
   
   \bibitem{gaplessCFL}
  Alford M, Kouvaris C, Rajagopal K.
  {\it Phys. Rev. Lett.} 92:222001 (2004)

\bibitem{Carlson:2005}
Carlson J, Reddy S, {\it Phys. Rev. Lett.} 95:060401 (2005)

   \bibitem{Huang:2004bg}
 Huang M, Shovkovy IA.
  {\it Phys. Rev. D} 70:051501 (2004)

  \bibitem{Casalbuoni:2004tb}
  Casalbuoni R, et al.
  {\it Phys. Lett. B} 605:362 (2005);
  erratum {\em ibid. B} 615:297 (2005)]

   \bibitem{nogapless2SC}
  Reddy S, Rupak G.
  {\it Phys. Rev. C} 71:025201 (2005);
   Giannakis I, Ren HC,
  {\it Phys. Lett. B} 611:137 (2005)


\bibitem{Casalbuoni:2005zp}
 Casalbuoni R, Gatto R, Ippolito N, Nardulli G, Ruggieri M,
{\it  Phys. Lett. B} 627:89 (2005)


\bibitem{SQM}
Itoh N.
{\it Prog. Theor. Phys.} 44:291 (1970);
Bodmer AR.
{\it Phys. Rev. D} 4:1601 (1971)
Witten E.
{\it Phys. Rev. D} 30:272 (1984)
Farhi E, Jaffe RL.
{\it Phys. Rev. D} 30:2379 (1984)

\bibitem{SQMReview}
Weber F,
  {\it Prog. Part. Nucl. Phys.} 54: 193 (2005)

\bibitem{StrangeCrust} 
Jaikumar P, Reddy S, Steiner AW.
{\it Phys. Rev. Lett.} 96: 041101 (2006)

\bibitem{Alcock}
Alcock C, Farhi E, Olinto AV.
{\it Astrophys. J} 310:261 (1986) 

\bibitem{Iwamoto}
Iwamoto N. {\it Ann. Phys.} 141:1 (1982)

\bibitem{qvisc}
Sawyer RF,
  {\it Phys.\ Rev.\ D} 39:3804 (1989)
  
  
\bibitem{CFLnuemis} 
Jaikumar P, Prakash M, Schafer T.
 {Phys. Rev. D} 66:063003 (2002);
 Reddy S, Sadzikowski M, Tachibana M.
 {\it Nucl. Phys. A} 714:337 (2003) 


\bibitem{CFLVisc}
Madsen J,
  {\it Phys. Rev. Lett.} 85, 10 (2000)

\bibitem{CFLK0nuemis} 
Reddy S, Sadzikowski M, Tachibana M.
{\it Phys. Rev. D} 68: 053010 (2003)

\bibitem{Alford:2004zr}
 Alford M, Jotwani P, Kouvaris C, Kundu J, Rajagopal K,
{\it Phys. Rev. D} 71:114011 (2005)


\bibitem{Pethick:1995di}
Pethick CJ, Ravenhall DG.
{\em Annu. Rev. Nucl. Part. Sci.} 45:429 (1995).


\bibitem{Glendenning:1992vb}
Glendenning NK.
{\em Phys. Rev. D} 46:1274 (1992)


\bibitem{pasta}
Ravenhall DG, Pethick CJ, Wilson JR,
{\em Phys. Rev. Lett.} 50:2066 (1983);
Ravenhall DG, Pethick CJ.
{\em Annu. Rev. Nucl. Part. Sci.} 45:429, (1995)

\bibitem{Heiselberg:1993dx}
Heiselberg H, Pethick CJ, Staubo EF.
{\em Phys. Rev. Lett.} 70:1355 (1993)

\bibitem{Glendenning:1998zx}
Glendenning NK, Schaffner-Bielich J.
{\em Phys. Rev. Lett.} 81:4564 (1998)

\bibitem{Norsen:2000wb}
Norsen T, Reddy S.
{\em Phys. Rev. C} 63:065804 (2001)

\bibitem{Voskresensky:2002hu}
Voskresensky DN, Yasuhira M, Tatsumi T.
{\it Nucl. Phys. A} 723:291 (2003)


%
%

 
\bibitem{Lattimer:2001} 
Lattimer JM, Prakash M.
{\it Astrophys. J.} 550:426 (2001)

\bibitem{Horowitz:2001}
Horowitz CJ, Piekarewicz J. 
{\it Phys. Rev. Lett.} 86:5647 (2001); 
{\it Phys. Rev. C.} 66:055803 (2002)

\bibitem{Li:2006}
Li B-A, Steiner AW. (2005) nucl-th/0511064.

\bibitem{Steiner:2005} 
Steiner AW, Prakash M, Lattimer JM, Ellis PJ.  
{\it Phys. Rep.} 411:325 (2005)

\bibitem{LPMY90}
Lattimer JM, Prakash M, Masak D, Yahil A
{\it Astrophys. J} 355:241 (1990)

\bibitem{WT03}
Weisberg JM, Taylor JH.
In {\em Radio Pulsars}, ed. M Bailes, DJ Nice, SE Thorsett, p. 93.
San Francisco: Astron. Soc. Pac., ASP Conf. Proc., Vol. 302
(2003) astro-ph/0211217

\bibitem{Yakovlev:2004}
Yakovlev DG, Pethick CJ.
{\em Annu. Rev. Astron. Astrophys.} 42:169 (2004)

\bibitem{Page1996}
Page D.
In {\em The Many Faces of Neutron Stars}, 
ed A Alpar, R Buccheri,  J van Paradijs, p. 539.
Dordrecht: Kluwer Academic Publishers (1998)

\bibitem{PA92}
Page D, Applegate JH.
{\it Astrophys. J} 394:L17 (1992)

\bibitem{PPLS00}
Page D, Prakash M, Lattimer JM, Steiner AW.
{\em Phys. Rev. Lett.} 85:2048 (2000)

\bibitem{Blaschke:2001}
Blaschke D, Grigorian H, Voskresensky DN.
{\em Astron. Astrophys.} 368:561 (2001)

\bibitem{PU:2002}
Page D, Usov VV.
{\em Phys. Rev. Lett.} 89:131101 (2002)

\bibitem{Gusakov}
Gusakov ME, Kaminker AD, Yakovlev DG, Gnedin OY.
{\em Astron. Astrophys.} 423:1063 (2004);
Kaminker AD, Gusakov ME, Yakovlev DG, Gnedin OY.
{\em Mon. Not. R. Astron. Soc.} 365:1300 (2006)

\bibitem{Blaschke:2004}
Blaschke D, Grigorian H, Voskresensky DN.
{\em Astron. Astrophys.} 424:979 (2004);0
Grigorian H, Voskresensky DN.
{\em Astron. Astrophys.} 444:913 (2005)

\bibitem{Rmode}
Andersson N, Kokkotas KD, Stergioulas N. {\it Astrophys. J} 516:307 (1999); 
Lindbolm L, Owen B. {\it Phys. Rev. D} 65:063006 (2002) 
 
\bibitem{Deepto}
Chakrabarty, D., in {\em Interacting Binaries: Accretion, Evolution, and Outcomes}  
Volume 797, pp. 71, AIP Conference Proceedings (2005)
 
 \bibitem{Madsen} 
 Madsen J. {\it Phys. Rev. Lett.} 81:3311 (1998) 
 
\bibitem{burr:86} Burrows A, Lattimer JM.
{\it Astrophys. J.}  307:178 (1986) 

\bibitem{Burrows:1992cy}
Burrows A, Klein D, Gandhi R,
{\it Nucl.\ Phys.\ Proc.\ Suppl.}  31:408 (1993).


\bibitem{SNnus}
Prakash M. In {\em The Nuclear Equation of State},
ed. A. Ansari, L Satpathy, pp. 229-410.
Singapore: World Sci. (1996);
Keil W, Janka HT.
{\it Astron. Astrophys.} 296:145 (1995);
Pons JA, et al. {\it Astrophys. J.} 513:780 (1999);
Pons JA, Steiner AW, Prakash M, Lattimer JM.
{\it Phys. Rev. Lett.} 86:5223 (2001);
Prakash M. et al. {\it Lect. Notes Phys.} 578: 364 (2001)

\bibitem{NuOpac}
Reddy S, In {\em Compact Stars: The Quest for New States of Dense Matter} ed. Hong, DK ; Lee, CH ; Lee, HK ; Min, DP ; Park, TS ; Rho, M, p. 495, Singapore: World Sci. ( 2003)

\bibitem{GW}
Cutler C, Thorne KS. 
In {\em General Relativity and Gravitation},
ed N Bishop, SD Maharaj, p. 72.
Singapore: World Scientific (2002) gr-qc/0204090

\bibitem{Chirp}
  Cutler C, Flanagan EE,
  {\it Phys.\ Rev.\ D} 49:2658 (1994)
  
\bibitem{Faber:2002zn}
  Faber JA, Grandclement P, Rasio FA, Taniguchi K.
  {\it Phys.\ Rev.\ Lett.} 89: 231102 (2002)
  
\bibitem{GWave}
  Miniutti G, Pons JA, Berti E, Gualtieri L, Ferrari V.
{\it Mon. Not. Roy. Astron. Soc.} 338:389 (2003)

\bibitem{SQMGwave}
Kluzniak W, Lee WH.
{\em Mon. Not. Roy. Astron. Soc.} 335:L29 (2002);
Ratkovic S, Prakash M, Lattimer JM.
arXiv:astro-ph/0512136 (2005) 

\end{thebibliography}
\end{document}